\documentclass{emulateapj}
\usepackage{apjfonts}
\usepackage{epsf}
\usepackage{psfig}
\begin{document}

\slugcomment{Submitted to ApJ}
\shortauthors{J. M. Miller et al.}
\shorttitle{Disks and Jets}

\title{On the Role of the Accretion Disk in Black Hole Disk--Jet Connections}

\author{J.~M.~Miller\altaffilmark{1},
  G.~G.~Pooley\altaffilmark{2},
  A.~C.~Fabian\altaffilmark{3}
  M.~A.~Nowak\altaffilmark{4},
  R.~C.~Reis\altaffilmark{1},
  E.~M.~Cackett\altaffilmark{3},
  K.~Pottschmidt\altaffilmark{5}$^{,}$\altaffilmark{6},      
  J.~ Wilms\altaffilmark{7}
}

\altaffiltext{1}{Department of Astronomy, University of Michigan, 500
Church Street, Ann Arbor, MI 48109, jonmm@umich.edu}
\altaffiltext{2}{Cavendish Laboratory, University of Cambridge, JJ
  Thomson Avenue, Cambridge CB3 OHE, UK}
\altaffiltext{3}{Institute of Astronomy, University of Cambridge,
Madingley Road, Cambridge CB3 OHA, UK}
\altaffiltext{4}{Kavli Institute for Astrophysics and Space Research,
MIT, 77 Massachusetts Avenue, Cambridge, MA 02139}
\altaffiltext{5}{CRESST and NASA Goddard Space Flight Center,
  Astrophysics Science Division, Code 661, Greenbelt, MD, 20771}
\altaffiltext{6}{Center for Space Science and Technology, University
  of Maryland Baltimore County, 1000 Hilltop Circle, Baltimore MD
  21250}
\altaffiltext{7}{Dr. Karl-Remeis-Sternwarte and Erlangen Center for
  Astroparticle Physics, Sternwartestr. 7, 96049, Bamberg, Germany}
\keywords{accretion, accretion disks, black hole physics, relativistic processes -- X-rays: binaries}

\begin{abstract}
Models of jet production in black hole systems suggest that the
properties of the accretion disk -- such as its mass accretion rate,
inner radius, and emergent magnetic field -- should drive and modulate
the production of relativistic jets.  Stellar-mass black holes in the
``low/hard'' state are an excellent laboratory in which to study
disk--jet connections, but few coordinated observations are made using
spectrometers that can incisively probe the inner disk.  We report on
a series of 20 {\it Suzaku} observations of Cygnus X-1 made in the
jet-producing low/hard state.  Contemporaneous radio monitoring was
done using the Arcminute MicroKelvin Array radio telescope.  Two
important and simple results are obtained: (1) the jet (as traced by
radio flux) does not appear to be modulated by changes in the inner
radius of the accretion disk; and (2) the jet is sensitive to disk
properties, including its flux, temperature, and ionization.  Some
more complex results may reveal aspects of a coupled disk--corona--jet
system.  A positive correlation between the {\it reflected} X-ray flux
and radio flux may represent specific support for a plasma ejection
model of the corona, wherein the base of a jet produces hard X-ray
emission.  Within the framework of the plasma ejection model, the
spectra suggest a jet base with $v/c \simeq 0.3$, or the escape
velocity for a vertical height of $z\simeq 20~GM/c^{2}$ above the
black hole.  The detailed results of X-ray disk continuum and
reflection modeling also suggest a height of $z \simeq
20~GM/c^{2}$ for hard X-ray production above a black hole, with a spin
in the range $0.6 \leq a \leq 0.99$.  This height agrees with X-ray
time lags recently found in Cygnus X-1.  The overall picture that
emerges from this study is broadly consistent with some jet--focused models
for black hole spectral energy distributions in which a relativistic
plasma is accelerated at $z=$10--100~$GM/c^{2}$.  We discuss these
results in the context of disk--jet connections across the black hole
mass scale.
\end{abstract}

\section{Introduction}
Numerous models have been proposed to explain how accretion can power
jets, and nearly all of them invoke magnetic fields at some level.
The ubiquity of collimated outflows in accreting systems is suggestive
of a common mechansism.  Magnetocentrifugal
acceleration, for instance, is one process that might work to launch
jets in settings as different as disks around young stars, and disks
around super-massive black holes (see, e.g., Blandford \& Payne 1982).
In this scenario, poloidal magnetic fields are anchored in the disk or
disk atmosphere, gas can escape along the field lines, and shocks
downstream in the flow can serve to further accelerate the material to
produce a relativistic jet.    

Evidence for magnetically--driven flows has been found in the
absorption spectra of young stars (e.g. Calvet, Hartmann, Kenyon
1993), in cataclysmic variable stars (Mauche \& Raymond 2000),
stellar-mass black holes (e.g. Miller et al.\ 2006, 2008; Kubota et
al. 2007), and also inferred in some active galactic nuclei (Kraemer et
al.\ 2006).  Imaging of the powerful relativistic jet in M87 on scales of
just $50~ GM/c^{2}$ shows a wide-angle outflow that is just starting
to collimate (Junor, Biretta, \& Livio 1999), and may futher support a
magnetocentrifugal origin.  Evidence for a helical magnetic field at
the base of a jet has recently been inferred from monitoring
observations of the ``blazar'' BL Lac (Marscher et al.\ 2008).

Another prevalent model for powering relativistic jets from black
holes invokes tapping the spin energy of the black hole via magnetic
field lines (Blandford \& Znajek 1977).  It is interesting to note
that the required field geometry is the same as the disk--driven
magnetocentrifugal case: poloidal field lines must be anchored in the
disk or disk atmosphere.  Some estimates of the work done on hot
cluster gas by powerful jets from super-massive black holes in the
largest central galaxies suggest that black hole spin must be getting
tapped (e.g. McNamara et al.\ 2009).  The dichotomy between radio-loud
and radio-quiet AGN can also be explained in terms of black hole spin
(Sikora, Stawarz, \& Lasota 2007), though the nature of the accretion
flow itself may also contribute to this dichotomy.  

Currently, there is no definitive evidence for or against black hole
spin contributing to jet power in stellar-mass black holes
(e.g. Fender, Gallo, \& Russell 2010; also see Miller et al.\ 2009,
Miller; Miller, \& Reynolds 2011; and Narayan \& McClintock 2012),
though it is clear that different methods of constraining spin are in
broad agreement and suggest high spin values in most systems (Miller
et al.\ 2009; McClintock et al.\ 2010; Miller, Miller,\& Reynolds
2011; Duro et al.\ 2011).  This situation is likely the result of some
obvious difficulties.  Theoretically, it is clear that a sort of
``throttle'' may act to tap the power of a spinning black hole.  This
throttle could be magnetic field strength or orientation, the mass
accretion rate, a combination or ratio of these quantities, or some
other physical quantity.  Observationally, difficulties inherent in
arranging coordinated X-ray and radio observations, limited angular
resolution for separating cores and knots in the radio band in the
Southern Hemisphere, and the frequent absence of gas bearing the
signatures of the work done by a jet all serve to complicate efforts
to reveal the role of spin.

Regardless of the details, changes in
the inner disk should produce changes in the jet.  The ``fundamental
plane'' of black hole accretion clearly demonstrates that jet
properties respond to inflow properties (Merloni, Heinz, \& DiMatteo
2003; Falcke, Kording, \& Markoff 2004; Gultekin et al.\ 2009; also
see King et al.\ 2011 and Jones et al.\ 2011).  One limitation of
these important relationships -- and, indeed, many joint X-ray and
radio studies of black holes -- is that X-ray monitoring observations
are seldom able to reveal properties of the {\em disk}.  This is
especially true for supermassive black holes because AGN disk emission
peaks in the UV (which suffers from extinction), and disk reflection
features can require relatively deep exposures (for a review, see Miller
2007; also see Nandra et al.\ 2007).  In constrast, the nature of the
disk in stellar-mass black holes can be constrained in numerous ways,
notably through its thermal emission (peaking in X-rays) and
reflection signatures.  However, shallow X-ray monitoring
observations, monitoring with coarse spectral resolution, and
monitoring confined to $E > 2-3$~keV cannot sample disk emission well.

For several reasons, Cygnus X-1 is an excellent source in which to
study the role of the accretion disk itself in launching jets.  First,
it is known to continuously launch relativistic jets in the
``low/hard'' state (Pooley, Fender, \& Brocksopp 1999; Stirling et
al.\ 2001; Wilms et al.\ 2006).  In this state, prior studies clearly
show that X-ray fluxes and radio fluxes from monitoring observations
are coupled (Gallo, Fender, Pooley 2003).  Radio flux modulations at
the level of 1-2 mJy are observed at the orbital period of
Cygnus X-1 (Pooley, Fender, \& Brocksopp 1999), but this amplitude is
small compared to the scale of the variability sampled in this
program.  Second, Cygnus X-1 is relatively close, with a distance of
1.86~kpc recently determined via radio parallax (Reid et al.\ 2011).
This ensures high X-ray and radio flux levels -- even in the low/hard
state -- and excellent spectra from soft X-rays up to very high energy
X-rays.  Last, spectra of Cygnus X-1 show two clear and independent
signatures of the accretion disk in the low/hard state: cool thermal
emission from the accretion disk, and relativistic reflection
including broad Fe K emission lines (e.g. Frontera et al.\ 2001;
Miller et al.\ 2002, 2006; Reis et al.\ 2010; Nowak et al.\ 2011; Duro
et al.\ 2011).

We therefore made a series of 20 observations of Cygnus X-1 with {\it
  Suzaku} in 2009.  The instruments aboard {\it Suzaku} provide
moderate-resolution CCD spectra in soft X-rays, coupled with extremely
sensitive and simultaneous hard X-ray coverage.  Thus, {\it Suzaku} is
able to detect the cool accretion disk, relativistic iron line, and
broad-band disk reflection spectrum in each observation.
Contemporaenously, we made frequent monitoring observations of the source
at 15~GHz using the updated Ryle radio telescope, now known as the
Arcminute Microkelvin Imager (AMI; Zwart et al.\ 2008).  

Comptonization models were fit to the HXD data obtained through this
program by Torii et al.\ (2011).  In this work, we report on efforts
to model the X-ray spectra using both phenomenological models
including simple disk continua and relativistic Fe K lines, and more
self-consistent physical disk reflection models.  The results of these
spectral fits -- particularly those tied to the disk -- were
correlated with radio flux to understand how disks influence jet
properties.  It should be emphasized that the central goal of this
program is to accurately characterize {\em relative} changes in disk
and jet properties, not absolute properties of the system.

\section{Observations Data Reduction}

\subsection{Radio Data}
The Arcminute MicroKelvin Imager arrays (AMI) are located in Cambridge, UK.
They consist of two aperture synthesis telescopes mainly devoted to
CMB-related studies (Zwart et al.\ 2008).  The observations described
in this paper were made with the Large Array (the reconfigured and
reequipped Ryle Telescope), consisting of eight 13~m antennas with a
maximum baseline of about 120~m, observing in the band 12--18 GHz.
The angular resolution is typically 25 arcsec.  Monitoring of
small-diameter sources is undertaken in the manner described by Pooley
\& Fender (1997) for the Ryle Telescope; observations are interleaved
with those of a phase-reference calibrator, and after appropriate
calibration the data for individual baselines are averaged as vectors.
The in-phase component then provides an unbiased estimate of the flux
density of the source in question.

The flux-density scale is calibrated by regular observations of 3C48,
3C147 and 3C286. The overall scale is believed to be consistent with
that used for the VLA (Perley, private comm.)  allowing for the use by
AMI of a single linear polarisation; all the AMI measurements are of
Stokes' I+Q.

Table 1 lists the start times and flux density at 15 GHz for 124 AMI
observations made across the time span of our {\it Suzaku}
observations.  These obesrvations are made on a best efforts basis,
and the observation spacing is sometimes irregular.  This small
downside is outweighed by the typical high cadence of these
observations and their sensitivity.  

In order to compare radio flux density to the X-ray spectral
properties derived in the fits described above, the start times of the
AMI and {\it Suzaku} observations were examined.  The AMI radio
observation most closely following each X-ray observation was
identified.  In this way, AMI observations were selected for direct
comparison to the 20 {\it Suzaku} observations.  This ordering is
based on the expectation that the X-ray-emitting inflow may help to
shape the radio jet, but that any influence of the radio jet on the
inflow is negligible by comparison.  

The mean delay between X-ray and radio observations in this program is
2.3 days.  This delay represents innumerable orbital timescales at the
innermost stable circular orbit around the black hole, but it is an
order of magnitude less than viscous timescales through the entire
accretion disk in X-ray binaries, which is expected to set the overall
character of the accretion flow geometry (e.g. Esin, McClintock, \&
Narayan 1997).

\subsection{Suzaku XIS Data}
Table 2 lists some basic parameters of the {\it Suzaku} observations
obtained through this program in 2009, including the XIS spectra
utilized from each observation.

Each {\it Suzaku}/XIS camera has a single CCD in a square array of
1024 by 1024 pixels.  The CCDs are 18 arcminutes on each side.  The
XIS0 and XIS3 cameras are front-illuminated CCDs, while XIS1 is back
illuminated.  The XIS2 camera is no longer functional due to a
micrometeoroid impact.  The breadth of the telescope PSF --
approximately 2.0 arcminutes (half-power diameter) -- is helpful in
spreading the flux of a bright incident source over many pixels.  In
addition, the XIS cameras are equipped with operational modes that can
further aid in the observation of bright sources.  The key issue is to
limit photon pile-up, which can falsely reduce the flux level that is
recorded and distort the spectrum by registering multiple incident
photons as single events.  The key quantity is the photon flux per
event box (typically 3$\times$3 pixel boxes used to select good X-ray events
and reject cosmic rays) per CCD frame time (see, e.g., Miller et
al.\ 2010).  Cygnus X-1 is a very bright source, and the instrumental
set-up was chosen to limit the effects of photon pile-up on the
resultant XIS spectra.

The HXD pointing position was used in all observations; in this
configuration, vignetting then reduces the effective area, lowering
the photon flux per event box.  In the vast majority of observations,
each camera was operated with a 1/4 window plus 0.5 second burst
option.  This has the effect of limiting the frame time, reducing the
photons registered in each event box per clock time, and again
limiting pile-up.  In a few cases, one of the cameras was run in
``PSUM'' mode, which offers very high time resolution at the expense
of imaging information.  PSUM mode is not robust against pile-up
effects for bright sources, and data taken in this mode are not
considered in this work.

All {\it Suzaku} data were reduced using version 6.9 of the HEASOFT
suite (including FTOOLS and CALDB).  The XIS data were reduced in
accordance with the {\it Suzaku} ABC Guide; these procedures and
additional conservative reduction steps are briefly described below:

XIS events can be recorded and telemetered in different editing modes;
2$\times$2 and 3$\times$3 are the most common, and 5$\times$5 is used less frequently.
Each unfiltered event file for each camera was reprocessed using the
tool ``xispi''.  New ``cleaned'' event files were then made using the
``xisrepro'' script, run within xselect.  This is often the full
extent of data processing for XIS data, but we took additional steps
to produce the best possible spectra:

$\bullet$ For each event file, we next ran the ``aeattcor.sl'' script
(Nowak 2009; Nowak et al.\ 2011), which corrects the data for the
large spacecraft wobble endemic to {\it Suzaku}.  This script
generates a new attitude file, which was subsequently applied to each
cleaned event list.  In every case, this procedure yielded a sharper
PSF.

$\bullet$ We then ran the XIS pile-up estimation tool (Nowak 2009;
Nowak et al.\ 2011), and produced images with contours corresponding
to different levels of photon pile-up.  Extraction regions were then
chosen so as to limit photon pile-up.  Spectra were extracted from
square boxes, 240 pixels on each side, with an inner exclusion circle
with a radius of 30 pixels and centered on the source coordinates.
This effectively limited photon pile-up in extracted spectra to 5\% or
less.  Backgrounds were extracted using the same region shape, offset
from the center of the PSF.

In this work, we only consider spectra obtained outside of the
well-known ``dipping'' intervals in Cygnus X-1.  As shown by
Balucinska-Church et al.\ (2000), dipping intervals vary with phase,
clustering around $\phi = 0$ (where the companion is closer to Earth).
This clustering suggests that the dips are partially due to
obscuration by inhomogeneities or ``blobs'' in the focused companion
wind.  Some dipping is seen at every orbital phase, however (again,
see Balucinska-Church et al.\ 2000), and thus all obserations must be
screened for dips.

Dipping intervals were identified and excluded using the methods
described by Nowak et al.\ (2011).  Briefly, light curves and
color-color diagrams were made using the XIS data, and intervals with
dip-like properties were identified.  These intervals were excluded by
specifying additional time selections in the standard good-time
interval (GTI) files.  This provides a relatively simple but effective
means of identifying and removing dipping intervals and the spectral
distortions that they introduce.

For each XIS camera, all event files were then loaded into ``xselect''
jointly.  Regions were defined according to the prescription above.
The standard goodtime filtering was applied.  Spectra, backgrounds,
final lightcurves, and final goodtime files were then extracted.

Redistribution matrix files were constructed for each spectrum using
the tool ``xisrmfgen'', and ancillary response files were made by
running the ``xissimarfgen'' tool with an input of 200,000 photons.
The XIS0 and XIS3 CCDs are both front-illuminated chips, with very
similar response functions.  In observations where both of these CCDs
were operated in imaging modes, their spectra, backgrounds, and
response functions were combined using the tool ``addascaspec''.

It is nominally the job of the redistibution matrix file to account
for over-sampling of the detector energy resolution, and to handle it
appropriately.  This file should be built based on a statistical
characterization of how often a photon - with an energy known a priori
- ends up in one of N channels that sample the relevant energy window.
In practice, this characterization is very difficult; the mutual
dependence of flux in different channels is hard to characterize
perfectly.  As a result, binning can tend to add more continuum into
weak and/or narrow line features than is nominally correct, and
thereby reduce the significance of such features.  Our goal is to
accurately characterize reflection in the Fe K band, which requires
separating its features from (weak and narrow) ionized absorption
features.  A stronger binning might legitimately reduce the
statistical uncertainty in some continuum and reflection parameters,
for instance, but might illegitimately reduce the complicating effects
of ionized absorption in the companion wind.  Although an even
stronger binning would better reflect the intrinsic energy resolution
of the XIS cameras, via the FTOOl "grppha" the XIS spectra were
grouped to have at least 10 counts per bin for the reasons cited
here.

Owing to its different response function, and its different (and
generally more severe) calibration uncertainties, XIS1 was excluded
from our analysis.  The single exception is the observation on 23 April, in
which both the XIS0 and XIS3 cameras were operated in ``PSUM'' mode.
The use of XIS1 in this case incurs a small but unavoidable systematic
uncertainty, relative to the other observations.

The XIS cameras suffer from calibration uncertainties at the edges of
their spectral bandpass, and in the Si region (see, e.g., Nowak et
al.\ 2011).  In order to avoid these problems, the XIS spectra were
fit in the 0.8-9.0~keV band.  In the front-illuminated XIS cameras,
the largest calibration issue in the Si band can be modeled with the
addition of a Gaussian absorption line at 1.86~keV.  (A negative
Gaussian centered at 2.44~keV is required to account for calibration
uncertainties in the XIS1 response function.)  In a small number of
cases, residuals in the Si band were markedly worse than in other
observations, with instrumetnal flux residuals extending up to
3.5~keV.  Nowak et al.\ (2011) noted similar problems; following that
work, the 1.5--3.5~keV band was ignored when fitting full relativistic
reflection models on May 25 and 29 and June 02 and 04.  These
observations do not occur at the same binary phase, nor do they occur
at particularly high or low flux levels.  This suggests a short-term
change in the instrument response rather than a change in the nature
of the spectra observed from Cygnus X-1.  In case this interpretation
is incorrect, the 1.5--3.5~keV range was included in fits with
phenomenological models.

\subsection{Suzaku HXD/pin Data}
The HXD/pin spectra were reduced in the standard manner recommended by
the ABC Guide.  This procedure is described here for clarity:

``Tuned'' background files for each observation were downloaded from
the {\it Suzaku} website.  These files take several weeks to
accumulate, and are available some time after an observation is made.
A common good time file was made by merging intervals from the cleaned
pin event file and the non-X-ray background event file.  Time filtering of
the cleaned event file was then performed within ``xselect'' using the
merged good time file, and a pin spectrum was extracted.  The
resulting spectral file was corrected for dead time using the tool
``hxddtcor'' and the appropriate housekeeping file.  

A non-X-ray background spectral file was also extracted using
``xselect'', after time filtering with the merged good time file.  As
recommended by the ABC guide, the exposure time of the non-X-ray
background spectral file was increased by a factor of 10 using the
ftool ``fparkey''.  To create the cosmic X-ray background spectral
file, the response file appropriate for the HXD pointing position and
gain and calibration epoch of each observation was selected from the
HXD calibration database.  This file was used to simulate a spectral
file within XSPEC, exactly as per the specifications in the ABC guide.
The non-X-ray background and cosmic X-ray background files were then
merged using the ftool ``mathpha''.

In all cases, the HXD/pin spectra were fit in the 12--70~keV band.
This range was found to be relatively free of strong calibration
uncertainties.  

\subsection{Suzaku HXD/GSO Data}
Source spectra and non-X-ray background spectra from the HXD/GSO were
generated and corrected for dead time using the ftool ``hxdgsoxbpi''.
The appropriate raw GSO background file for each observation was
downloaded from the public archive.  The specral response files
(including the ``correction arf'') appropiate for each observation
were also downloaded from the public archive.

In all cases, the HXD/GSO spectra were fit in the 70--500~keV band.
This range was found to be relatively free of strong calibration
uncertainties.  The upper limit of 500~keV is the highest energy at
which the source is detected in all of the observations.

\subsection{Correcting For Ionized Absorption}
In order to account for the effects of X-ray absorption in the
companion wind, we downloaded and analyzed the longest {\it
  Chandra}/HETG observation in the ``low/hard'' state available in the
public archive.  Specifically, we analyzed a 55.85~ksec expsosure made
starting 2003 March 04 at 15:45:02 UT (sequence 400302, obsid 3815).
To prevent photon pile-up, the ACIS-S array was read-out in
``continuous clocking'' mode, and a gray filter was used on the S3
chip to prevent frame dropping.  (For details and additional examples
of this instrumental set-up, see Miller et al.\ 2006; 2008).

The data were obtained using the {\it Chandra} ``TGCAT'' facility,
which archives calibrated, up-to-date, level-2 event files, spectral
files, and redistribution matrix files.  All processing was performed
using CIAO 4.2.  Using the tool ``dmtype2split'', we divided the
delivered ``pha2'' file into individual components.  Ancillary
response files were generated using the tool ``mkgarf'' and combined
the first-order MEG spectra and HEG spectra.  These spectra were then
analyzed using XSPEC.

As reported in many prior papers, the combined MEG and HEG spectra of
Cygnus X-1 contain absorption lines that originate in the companion
wind.  The wind is not thought to be point symmetric nor even
axisymmetric.  Rather, comparisons from different points in the binary
phase suggest that the wind is ``focused'' (e.g. Sowers et al.\ 1998;
Miller et al.\ 2005; Gies et al.\ 2008; Hanke et al.\ 2009). The
column density in the wind is maximal when the companion star is
closest to the observer ($\phi = 0$), and is reduced when the black
hole primary is closer to the observer ($\phi = 0.5$).

At present, no systematic, self-consistent photoionization study of
spectra obtained at different binary phase points has been performed.
Indeed, treatments in the literature vary widely (though the reader is
refered to Hanke et al.\ 2009 as a thorough treatment).  To address
the absorption observed in this long {\it Chandra} observation, we
constructed a large grid of XSTAR models using ``xstar2xspec''
(version 2.2).  The SED used to create the XSTAR grid was taken from
disk blackbody plus power-law model ($kT=0.19$~keV, $\Gamma = 1.8$,
commensurate with values found in fits to the {\it Suzaku} data) used
to describe the {\it Chandra} continuum.   This grid was then included
in direct fits to the MEG and HEG spectra as a table model that
multiplied a simple disk blackbody plus power-law continuum.  In these
fits, the ionization parameter of the absorbing gas, its equivalent
neutral hydrogen column density, and its velocity shift were all free
parameters.

Based on the ephemeris given by Brocksopp et al.\ (1999), this long
{\it Chandra} observation started at a binary phase of $\phi = 0.77$.
Direct fitting suggests a number density of ${\rm log}(n) = 11$, an
ionization parameter of ${\rm log}(\xi) \simeq 2$ (where $\xi =
L/nr^{2}$), and an equivalent neutral hydrogen column density of ${\rm
  N}_{\rm H} \simeq 4\times 10^{21}~ {\rm cm}^{-2}$, and a modest
red-shift give a reasonable description of the data (the strongest
lines from abundant elements are fit well).  This is in broad
agreement with other results obtained near to this phase (Miller et
al.\ 2005; Hanke et al.\ 2009).  In more detailed treatments,
individual ions may require slightly different parameters, However,
these values are representative of the results obtained even as the
column density varies with binary phase (see Hanke et al.\ 2009).
Last, it is important to note that this model correctly accounts for
Fe XXV and Fe XXVI absorption lines that fall on top of the
relativistic disk line in Cygnus X-1 (see Figure 1).

In all fits to the {\it Suzaku} spectra, then, this XSTAR table model
was included.  Although ionized absorption lines are evident in the Fe
K band after fitting for disk reflection, the resolution of the XIS
makes it difficult to clearly detect weak lines at lower energy, at
least at the level required to constrain the ionization of a global
model.  Therefore, an ionization parameter of ${\rm log}(\xi) = 2.0$
was held fixed when fitting the XSTAR table model.  Moreover, the
red-shift of ${\rm few}\times 100$~km/s found in fits to the {\it
  Chandra}/HETG spectra is insignificant at the resolution of the XIS,
so a velocity shift of zero was used in all cases.  Thus, only the
column density was allowed to vary.

Whether or not CCD spectra could usefully constrain the parameters of
an {\it ionized} absorption zone was an open question.  To this end,
the column density measured via phenomenological X-ray spectral fits
(see Section 3.1) is plotted versus binary phase in Figure 2.  The
expected variation is evident in the data: the column density is
highest close to $\phi = 0$, and reduced close to $\phi = 0.5$.  This
is a strong indication that our modeling is able to correctly account
for variations in the spectrum due to the companion wind, and to
differentiate those variations from true changes in the low-energy
disk flux and relativistic line flux.  We remind the reader that
dipping intervals were excluded from the spectra in this exercise and
in the analysis that follows.

\section{X-ray Spectral Models}
In all fits, the equivalent neutral hydrogen column density was fixed
to ${\rm N}_{\rm H} = 6\times 10^{21}~ {\rm cm}^{-2}$, as per the
results of fits to high resolution {\it Chandra}/HETG spectra (Schulz
et al.\ 2002).  Although the ionized column density in the companion
wind may change as a function of orbital phase, the neutral column
density is likely best modeled as being constant (Miller, Cackett, \&
Reis 2009), at least in non-dip spectra and at phases other than
$\phi=0$.  The ``tbabs'' model within XSPEC was used to fit the
interstellar column density (Wilms, Allen, \& McCray 2000), assuming
the elemental abundances given by Anders \& Grevesse (1989).

As noted above, {\it all} spectral models were additionally modified
by the photoionized absorber described above.  The ionization
parameter was fixed at log$(\xi) = 2.0$, but the equivalent hydrogen
column density of this {\it ionized} absorption was allowed to vary in
phenomenological fits to each observation.  For simplicity and
expedience, the equivalent hydrogen column density measured in each
observation using phenomenological models was then held fixed when fitting
the more physical reflection models.

\subsection{Phenomenological Spectral Models}
Figure 3 shows a typical broad-band spectrum and the associated
data/model ratio that results from fitting with a simple power-law
model.  The spectrum was fit by ignoring the spectrum below 3~keV, in
the Fe K band (taken to be 5--7~keV here, based on typical broad
emission line profiles), and in the 15-45 keV band.  The strong soft
excess is evidence of thermal emission from an accretion disk, and the
curvature seen through the HXD bandpass is strongly indicative of disk
reflection.

Disk reflection models can differ in their assumptions.  Some are
especially suited to high ionization (e.g. ``reflionx'', see Ross \&
Fabian 2005; also see Dovciak, Karas, \& Yaqoob 2004), some are suited
to low ionization (e.g. ``pexrav'', see Magdziarz \& Zdziarski 1995).
Most models assume a constant density in the upper levels of the disk
where reflection occurs; this might be appropriate if e.g. magnetic
pressure dominates.  Other models assume a vertical density gradient
determined by hydrostatic equilibrium (e.g. ``xion'', see Nayakshin \&
Kallman 2001).

In order to provide relatively model--independent results, then, we
initially fit every observation with a phenomenological spectral
model.  The model effectively replicates a relativistically-blurred
disk reflection spectrum with a simple continuum, but it is easily
reproduced and it is free of assumptions regarding disk structure, the
nature of the interaction between the corona and disk, and the
physical processes at work in the corona.  (The results of these fits
can be compared to the Comptonization fits made to the same HXD data
by Torii et al.\ 2011.)

The soft excess at low energy was fit using a disk blackbody model
with a modified emissivity law (the ``diskpbb'' model in XSPEC;
e.g. Mineshige et al.\ 1994).  This is a generalization of the
better-known ``diskbb'' model (Mitsuda et al.\ 1984), in that the disk
temperature is assumed to vary with radius according to $T \propto
r^{-p}$ far from the black hole.  The ``diskbb'' model assumes the
``standard'' value of $p = 3/4$.  In light of recent results showing
that disks may be strongly irradiated by the corona in the low/hard
state (Rykoff et al.\ 2007; Gierlinski, Done, \& Page 2009; Reynolds
et al.\ 2010; Reynolds \& Miller 2012), we have assumed $p=0.5$ as per
an irradiated accretion disk.  This does not bias the disk radius to
be close to the black hole; rather, more of the flux is assumed to
originate from larger radii.

Again, the focus of this paper is on how parameters {\it vary}, not
absolute values.  However, the disk continuum does permit constraints
on the inner disk radii -- often a parameter of interest -- and we
have derived radii based on the continuum as self-consistently as
possible.  The flux normalization of ``diskbb'' and ``diskpbb''
components is given by $K = (r/\mathrm{km})^{2}~ \cos\theta
(d/10\,\mathrm{kpc})^{-2}$.  In calculating radii in gravitational
units, we have assumed $d = 1.86$~kpc (Reid et al.\ 2011), an
inclination of $\theta = 27^{\circ}$, and a mass of $M =
14.8~M_{\odot}$ (Orosz et al.\ 2011).  This inclination is that of the
binary system; the inner disk may be misaligned (for a discussion, see
Maccarone 2002), and we are simply assuming that the binary and inner
disk inclinations are consistent.  Given that
gravitomagnetohydrodynamics tends to anchor the inner disk
perpindicular to the black hole spin vector (Bardeen \& Petterson
1975), and that jet production should be aligned with that vector, the
absence of a two--sided jet in radio images of Cygnus X-1 (Stirling et
al.\ 2001) suggests that the inner disk is also viewed at a low
inclination.

The ``diskbb'' model is a standard against which correction factors
and calibrations have been calculated.  In this analysis, the
``true'' inner disk radius is calculated via $r_{in} = \eta f_{gr}^{2}
f_{col}^{2} r_{col}$/$\sqrt(g)$, where $\eta$ is the ratio of the
inner disk radius to the radius at which the emissivity actually
peaks, $f_{gr}$ accounts for relativistic effects, and $g$ accounts
for relativistic effects on apparent inclination (we have assumed
$\eta = 0.7$, $f_{gr} = 0.5$, and $g = 0.3$, as per Zhang, Cui, \&
Chen 1997, with the assumption that Cygnus X-1 likely has a high
spin).  The color correction factor, $f_{col}$, accounts for radiative
transfer (we have assumed $f = 1.7$ as per Shimura \& Takahara 1995
and Merloni, Ross, \& Fabian 2000).  The value of $\eta$ that we have
taken is likely to be marginally too high, in that it assumes a disk
where $T \propto r^{-3/4}$ holds at large radii; a disk following $T
\propto r^{-1/2}$ will dissipate most of its energy at even larger
radii and would nominally require a smaller value of $\eta$.

The medium and high energy spectrum was fit using a broken power-law
component (``bknpow''), modified by an exponential cut-off at high
energy (``highecut''; see Wilms et al.\ 2006 for background on this
simple model for the high energy spectrum of Cygnus X-1).

The Fe K band in Cygnus X-1 is known to be comprised of at least two
components: a narrow neutral emission line arising either though
irradiation of clumps in the companion wind or the outer disk, and a
broad relativistic line that is dynamically broadened in the innermost
accretion disk (see, e.g. Miller et al.\ 2002, Miller et al.\ 2005,
Nowak et al.\ 2011, Duro et al.\ 2011).  The narrow, neutral line is
very narrow even at HETG resolution (see Figure 1), so it was modeled
using a zero-width Gaussian component with a centroid energy frozen at
6.40 keV.  Although early attempts to constrain the spin parameter in
Cygnus X-1 yielded inconsistent results (e.g., Miller et al.\ 2005;
Miller et al.\ 2009), more recent efforts using improved models and
data have converged on a high spin (Steiner et al.\ 2011; Duro et
al.\ 2011).  The broad line was therefore modeled using the ``Laor''
line model (Laor 1991).  A drawback of the Laor model is that it
assumes maximal spin; however, it allows a large range of disk radii
to be explored, and the focues of this work is to understand relative
rather than absolute values.  The line centroid energy was bounded
between 6.70--6.97~keV (Fe XXV--XXVI) as per an ionized disk.  The
emissivity index was fixed at $q=3$, (where ${\rm J(r)} \propto {\rm
  r}^{-q}$); this was motivated by preliminary phenomenological and
reflection fits that all returned values of $q$ consistent with 3.0.
The inclination was fixed at $27^{\circ}$, again assuming that the
black hole spin vector is aligned with the binary inclination (but see
Maccarone 2002).  The outer illumination radius was fixed at $400~
{\rm GM}/{\rm c}^{2}$, and the inner radius was allowed to vary.

The total phenomenological spectral model used was:\\
const$\times$tbabs$\times$mtable\{wind\_abs\}$\times$
(gauss+laor+diskpbb+bknpow)$\times$highecut.

\subsection{Relativistically-Blurred Reflection Models}
To provide a more physical description of the inner accretion flow, we
also fitted each of the broad-band spectra with the ``reflionx'' disk
reflection model (Ross \& Fabian 2005; note that Comptonization
models, sometimes invoking multiple Comptonization zones, may also
provide a reasonable physical description of the data; see Torii et al.\ 2011).
This model assumes a constant density for the accretion disk, and the
emergent spectrum is averaged over viewing angle.  However, it is
suited to a broad range of ionization parameters; it includes emission
lines from C, N, O, Ne, Mg, Si, S, and Fe; and its temperature and
ionization calculations do not assume local thermodynamic equilibrium.
Unlike some reflection models, ``reflionx'' is calculated over the 1
eV to 1 MeV range, so it can safely be applied across the entire
0.8--500~keV bandpass of our {\it Suzaku} spectra.  The broad range of
elements and charge states considered in ``reflionx'' is of importance
at low energy, where emission lines from reflection may blur together
to create a pseudo-continuum.  Fits with ``reflionx'', then, are
crucial to resolving how much of the apparent disk emission is
actually due to viscous dissipation in the disk, and how much might be
due to reflection.

% Important to highlight this in the results section

In fitting ``reflionx'', the iron abundance was fixed at the default
(solar) value, and the source red-shift was fixed at zero.  The photon
index of the power-law assumed to irradiate the disk within
``reflionx'' was linked to the photon index of a separate, simple,
unbroken power-law continuum component.  The ionization parameter of
the accretion disk and the flux normalization of the ``reflionx''
component were both allowed to float freely.  The reflection spectrum
was blurred using the ``kdblur'' model, which is based on the Laor
line function.  As per the phenomenological model, the inclination was
fixed at $27^{\circ}$, the emissivity was fixed at $q=3$, the outer
illumination radius was fixed at $400~{\rm GM}/{\rm c}^{2}$, and the
inner disk radius was allowed to float freely.  The narrow Fe K
emission line at 6.4~keV was included as before.  The soft thermal
emission was again modeled using a generalized disk blackbody assuming
irradiation (``diskpbb'' with $p=0.5$), and the total spectral model
was modified by a high energy cut-off (``highecut'').  The total
spectral model used was:

const$\times$tbabs$\times$mtable\{wind\_abs\}$\times$ (gauss+dispkbb+pow+kdblur$\times$atable\{reflionx.mod\}) $\times$highecut.

It must be noted that the total spectral model is not entirely
self-consistent.  The ``reflionx'' model assumes an exponential
cut-off with an e-folding energy fixed at 300~keV.  This is generally
a factor of two higher than is measured in fits with the
phenomenological model (see below).  Owing to this disparity,
acceptable reflection fits cannot be obtained unless a separate
cut-off is applied.  This could have the effect of accounting for some
curvature that is properly due to reflection instead through a
spectral cut-off.  The reflection results are driven by the features
and sensitivity achieved below 10 keV, and then again in the
20--40~keV range (where the Compton back-scattering ``hump'' peaks),
and less by the nature of the spectrum at and above 300~keV.

It should also be noted that ``reflionx'' does not measure the
``reflection fraction'', unlike some prior reflection models.  This is
partly because there is no longer a simple linear relationship between
the incident and reflected flux, when ionization is included and
properly treated.  Rather, the albedo of the disk {\it increases} with
ionization at high ionization parameters.  This is even more true when
the ionization balance within the disk is actually being solved by the
reflection model.  Moreover, the values obtained in calculating a
``reflection fraction'' are strongly dependent upon the energy band
considered, and values obtained via blurred ionized reflection models
fit over the 0.8--500.0 keV band would not easily map to values
obtained with older, un-blurred models fit to a standard RXTE bandpass
(e.g. 3-100 keV).

Last, the ``Laor'' relativistic line model (Laor 1991), and the
``kdblur'' convolution model based on ``Laor'', enable relatively
quick spectral fits and are sufficient to characterize relative
differences in line properties.  Newer line models that include spin
as a variable parameter are now available (e.g. Dovciak, Karas, \&
Yaqoob 2004, Beckwith \& Done 2004, Brenneman \& Reynolds 2006).
Although such models might be better suited to measuring absolute
values, they can greatly increase the time required to make a fit.
New line models appear to give improvements over older models at the
10\% level (Beckwith \& Done 2004).

\section{X-ray Spectral Fitting Results}
\subsection{Initial Checks}
Prior to interpreting the results of X-ray spectral fits, and the
implications of correlations between X-ray fitting parameters and
radio flux density, the X-ray and radio data were checked for
consistency with prior results in a model--independent fashion.  The
HXD/pin count rate was selected for comparison, to best approximate
the kind of data that are obtained in typical X-ray monitoring
observations.  Figure 4 shows light curves of the radio flux density
and HXD/pin count rates; the data were divided by their mean values in
order to show the relative strength of flux variations.  In order to
put the X-ray observations in a broader context, the full set of 124
radio observations listed in Table 1 are plotted in Figure 4.  It is
clear from this figure that the radio and hard X-ray points track each
other well, and that the radio flux shows higher fractional
variability than the X-ray flux.

Figure 5 plots radio flux density versus the HXD/pin count rate, for
the 20 {\it Suzaku} observations and the radio observations that
followed most closely after each.  The quantities are clearly
correlated, though there is scatter.  A Spearman's rank correlation
test gives coefficient of $\rho = 0.45$, corresponding to a 4.4\%
chance of a false correlation.  Thus, the {\it Suzaku} and AMI data
confirm the radio--X-ray coupling that has been seen previously with
X-ray monitor data (see Gallo, Fender, \& Pooley 2003, Wilms et
al.\ 2006, and Zdziarski et al.\ 2011).  This provides a measure of
confidence that our program has sampled Cygnus X-1 in a period of
typical disk--jet coupling.  The prior demonstratons of a postive
radio and X-ray flux correlation also show a degree of scatter,
despite the higher cadence of X-ray flux points provided by the
RXTE/ASM.  This suggests that the scatter in Figure 5 is not
exclusively due to a lack of strict simultaneity.

\subsection{Basic Results}
The results detailed in Tables 3 and 4 are generally good fits to the
spectra, with $\chi^{2}/\nu \simeq 1$.  Figure 6 depicts a typical fit
to the broad-band spectrum with the phenomenological model.  Figures 7
and 8 show the amplitude of the broad and narrow Fe K lines, when the
data are fit with a phenomenological continuum.  Figure 9 shows the a
typical fit to the broad-band spectrum of Cygnus X-1 with the more
physical relativistically--blurred disk reflection model.

Where the fits are not formally acceptable, there are likely three
main causes.  First, calibration uncertainties produced especially
poor fits to the specta obtained on certain days, for instance on 29
May.  Second, the ``reflionx'' reflection model has a hard lower bound
of $\Gamma = 1.4$ for the power-law index of the hard emission assumed
to illuminate the disk.  A number of fits pegged at this limit,
suggesting that a slightly harder power-law might have been favored by
the data.  Last, the hard emission may be better described using a
broken power-law, and ``reflionx'' only assumes a simple power-law at low energy.
It is notable that most spectra are fit better with the
phenomenological model than with the blurred reflection model.  Other
efforts have also found that physical models do not provide better
fits to spectra from Cygnus X-1 than phenomenological models (Wilms et
al.\ 2006; Nowak et al.\ 2011).  This is likely the result of the
greater freedom afforded by a broken power-law and separate line
component, and the limitations of ``reflionx'' noted above. Overall,
the quality of the spectral fits -- both phenomenological and physical
-- is such that the resulting measurements and correlations are
meaningful.

Given the complexities of these observations, including the use of the
HXD pointing pointing position for the XIS, the spacecraft wobble,
attempts to correct for the wobble, the use of window and burst
options, efforts to mitigate photon pile-up in the XIS, and evolution
of the HXD performance, it is not clear what the detector
cross-normalization factors (captured by the ``constants'' in both
families of spectral models) should be.  The use of constant factors
to normalize across different detectors when fitting over a broad
energy range is well established (see, e.g. Sobczak et al.\ 2000,
concerning multiple spectra of XTE J1550$-$564 obtained with RXTE).
Moreover, some of the relevant effects likely change between
obsevations.  The constant factors were therefore allowed to float in
each observation.  Values for the constant fell in the 0.90--1.10
range in the phenomenological fits, and in the 1.02--1.32 range for
the blurred reflection fits (values for the PIN/GSO factor fell in the
0.75--1.05 range, and 0.80-1.17 range, respectively).  A similar
observational set-up was used to observe the black hole candidate MAXI
J1836$-$194 using {\it Suzaku} (Reis et al.\ 2012).  The values of the
cross-normalization factors found in our fits are broadly consistent
with values covered by the error contours derived from fits to the
spectra of MAXI J1836$-$194.  Small changes to the factors we recorded
(e.g. 0.01) cause small changes in the continuum fits, not the line
properties.  Thus, although uncertainty in the flux offsets could be
assessed as a systematic error, it is not one that has a strong effect
on the featues of interest.  Larger changes (e.g. 0.1), or setting the
XIS/PIN factor to a value of 1.17 (common in more standard modes),
results in dramatically worse fits in many cases, such that no
phenomenological nor physical spectral model is acceptable.

In every sense, the phenomenological X-ray spectral fitting results
are typical of Cygnus X-1 in the ``low/hard'' state.  The power-law
index above the spectral break is found to be quite hard, generally in
the range of $1.4 \leq \Gamma_2 \leq 1.5$.  Similar indices are
measured via fits with the blurred ``reflionx'' model.  The cut-off
energy measured in the phenomenological fits is generally consistent
with $kT = 23$~keV.  This could be indicative of the electron
temperature of the hard X-ray corona, but it is likely that much of
the curvature fit by this low energy cut-off is actually due to disk
reflection.  This possibility is supported by the fact that $\Gamma_2$
is found to be close to the simple power-law in the reflection fits
(see Table 3 and Table 4), and by the fact that the power-law break
energy is generally found to be at or above the Fe K line energy,
where reflection is expected to produce a similar change in the
continuum.

In both families of fits, the temperature of the disk component is
found to be quite low, generally at or below $kT \simeq 0.2$~keV.
Again, this is fully consistent with black hole X-ray binaries in the
``low/hard'' state generally, and Cygnus X-1 in particular (see, e.g.,
Reis, Fabian, \& Miller 2010; Reynolds \& Miller 2012).  The
temperature values lie below the energy boundary considered in fits to
the XIS spectra; however, simple blackbody functions peak at $2.8~
kT$, and disk blackbody functions peak at a slightly higher energy.
Thus, although the accretion disk is found to be cool in all cases, a
good deal of its flux falls within the XIS fitting range and serves to
constrain the fitting parameters.  Disks in this temperature and flux
range cannot be detected in observations obtained with the RXTE/PCA
and INTEGRAL/JEM-X cameras, for instance, upon which prior coordinated
observing campaigns have been built (Wilms et al.\ 2006).

The inner disk radius implied by the disk continuum is of interest.
Tables 3 and 4 give the normalization of the disk component, and the
radius implied for $d = 1.86$~kpc (Reid et al. 2011), $M = 14.8~
M_{\odot}$, and $i = 27^{\circ}$ (Orosz et al.\ 2011), as well as
various correction factors.  Although a couple of observations imply
greater deviations, most normalizations differ by less than a factor
of $\leq 2$, which would correspond to radii generally varying by
less than 40\%.  It is immediately apparent that this modest variation
is far less than the factor of $\sim4$ variation seen in the radio
flux in Figure 4.  The normalizations of the disk components are
generally higher in the relativistic reflection fits ($r_{mean} = 3.9~
GM/c^{2}$, $\sigma_{r} = 0.6~ GM/c^{2}$) than in the phenomenological
fits ($r_{mean} = 3.4~ GM/c^{2}$, $\sigma_{r} = 0.8~ GM/c^{2}$),
suggesting that a small part of the putative disk continuum could
potentially be due to blurred atomic lines from disk reflection.
Including all of the relevant correction factors, none of the 40 disk
continuum fits imply a radius greater than $6~GM/c^{2}$.  Indeed, 33
of 40 disk continuum fits suggest $r_{in} \leq 4 GM/c^{2}$, which
implies a spin parameter of $a \geq 0.6$.

Fits to the relativistic line and blurred reflection spectrum provide
an independent check on the innermost extent of the accretion disk,
and variations in that radius.  Phenomenological fits with a separate
Laor line component generally find radii between 3--4$~GM/c^{2}$
($r_{mean} = 3.5~ GM/c^{2}$, $\sigma_{r} = 0.6~ GM/c^{2}$), in good
agreement with fits to the continuum.  All radii inferred using fits
with the additive ``Laor'' line component are found to be
significantly below $6~GM/c^{2}$.  Radii measured via the blurred
reflection fits give a broader range of values ($r_{mean} = 4.7~
GM/c^{2}$, $\sigma_{r} = 5.2~ GM/c^{2}$), and less agreement with the
disk continuum.  A number of fits with the relativistically-blurred
reflection component imply a radius only marginally larger than
$1.2~GM/c^{2}$, corresponding to maximal black hole spin.  A small
number of fits imply a larger disk radius, e.g. $r = 20\pm 9~
GM/c^{2}$ on 06 May.  Observations in which larger radii are found
generally represent poor measurements: if only those observations with
${error}(r)/r \leq 3$ are considered, all radii are below $6~GM/c^{2}$
and values approach maximal spin.  Indeed, weighting the values by the
inverse of their errors gives a mean radius of $1.7~ GM/c^{2}$.
Discarding poor measurements, and assuming that the smallest measured
values indicate the actual innermost radius defined by the spin of the
black hole, the disk continuum and disk reflection fitting results
point to a spin in the range $0.6 \leq a \leq 0.99$.

\subsection{Tests of X-ray Disk Reflection}

This program represents the first time that a CCD spectrometer has
observed a single source on 20 occasions, each time achieving the
sensitivity needed to detect a relativistic line and disk reflection
spectrum.  Therefore, some simple checks on the choice of a reflection
geometry are very much in order.  For instance, simple reflection
models predict that the flux of a relativistic line should follow the
flux of the power-law continuum that illuminates the disk (e.g. George
\& Fabian 1991).  Figure 10 shows a tight correlation between the flux
in the Laor line component and power-law component in phenomenological
fits.  A Spearman's rank correlation test gives a coefficient of
0.958, or a probability of just $3.3\times 10^{-11}$ of a false
correlation.  This may represent clear suppport for the illuminated
disk geometry assumed by simple disk reflection models.

Evidence of feedback between the disk and corona is predicted to take
the form of a relationship between the fraction of the incident flux
that is reflected by the disk, and the incident power-law photon index
(the $R-\Gamma$ correlation; Gilfanov, Churazov, \& Revnivtsev 1999;
Zdziarski et al.\ 2003).  Evidence of this correlation has been seen
in fits to gas spectrometer data (e.g.Gilfanov, Churazov, \&
Revnivtsev 1999; Zdziarski et al.\ 2003), but this program represents
the first opportunity to test the correlation using CCD spectra in
which relativistic lines are better revealed and separated from narrow
components.  

In the case of neutral reflection, the equivalent width of the
relativistic line can serve as a proxy for the reflection fraction as
the two are closely related (George \& Fabian 1991 predict that $R
\sim EW/180~{\rm eV}$, where $R$ is the reflection fraction).  Figure
11 plots the equivalent width of the Laor line component versus the
soft power-law index measured in phenomenological fits to the X-ray
spectra.  A tight, positive correlation is again observed.  A
Spearman's rank correlation test gives a coefficient of $\rho =
0.921$, with a false correlation probability of just $3.9\times
10^{-9}$.  The slope of this relationship is clearly greater than
unity, and broadly consistent with $\simeq 2$.  This is broadly
consistent with the linear relationship expected between these
parameters, and thus broadly consistent with feedback between the disk
and corona.  The fact that the slope is greater than unity may hint at
the need to treat reflection more self-consistently, changes to the
disk albedo due to ionization, and/or the influence of graviational
light bending, which predicts a slope steeper than unity if the hard
X-ray source is very close to a spinning black hole (Miniutti \& Fabian
2004).

Spectral fits with the relativistically-blurred disk reflection
spectrum allow the flux in the directly-observed power-law and the
reflection component to be measured independently and
self-consistently.  It is interesting to note that the reflected flux
-- not the direct power-law flux -- is found to be the {\it dominant}
flux component in half of the observations (see Table 4).  Moreover,
taking the ratio of the reflected flux to the total flux as a proxy
for the reflection fraction $R$, there is no significant correlation
between $R$ and $\Gamma$ in the full reflection fits.  Together, these
results suggest that feedback between the disk and corona, and perhaps
the nature of the corona itself, may be more complex than prior studies
revealed.  

Yet it is difficult to discount the extremely tight correlation
between the relativistic line flux and power-law flux (see Figure 10
and the correlations listed in Table 6) based on the phenomenological
fits.  That correlation is so tight that it calls into doubt the
results obtaind from the more physical models including a blurred
reflection component (see below).  There is a simple way in which
these apparently contradictory results can be reconciled, however: the
power-law in the phenomenological fits is a {\it broken} power-law,
meant to mimic a proper reflection spectrum.  The flux in this
component may not be direct flux, but rather a mixture of direct and
reflected flux.  Correlating the line and continuum flux would then be
a spurious procedure of (at least partly) correlating a component
against itself.  A very tight relationship {\it should} result from
such a test.  More discussion of the reflection models and their
consequences follows in Section 6.

Fits with the relativistically-blurred ``reflionx'' component also
suggest a moderately high ionization parameter, generally in the $3 <
{\rm log}(\xi) < 4$ range (see Table 4).  As with the other parameters
measured in fits to these {\it Suzaku} observations, this level of
ioniziation is typical of the low/hard state (see, e.g., Reis et
al.\ 2008, Blum et al.\ 2009).  In this regime, He-like and H-like
charge states of Fe are expected to be the most prominent.  The Laor
line centroid energy values were restricted to the range appropriate
for these charge states, but many of the fits measured centroids
within this range.  Thus, the reflection
fits and the separate line fits are in broad agreement with regard to
the ionization of the disk.

The details of fits to the narrow Gaussian line fixed at 6.40~keV are
omitted from Table 3 and Table 4 owing to space limitations, and
because this feature is not of central importance.  The narrow
emission line flux was generally found to be consistent with
$1.0\times 10^{-3}~ {\rm ph}~ {\rm cm}^{-2}~ {\rm s}^{-1}$ and an
equivalent width of 15--20~eV.  These values are consistent with prior
detections of this line component with ASCA (Ebisawa et al.\ 1996) and
at high resolution using the {\it Chandra}/HETGS (Miller et al.\ 2002;
Nowak et al.\ 2011).  Moreover, these values are consistent with the
illumination of cold clumps in the companion wind or the outer
accretion disk (Ebisawa et al.\ 1996).

\section{Correlating X-ray and Radio Parameters}
The results of running Spearman's rank correlation tests on different
pairings of radio flux density and X-ray spectral fitting parameters,
and between different pairings of X-ray spectral fit parameters only,
are given in Tables 5--8.  The detailed results of these tests are
given below.

\subsection{On Disk Radius and Jet Power}
Figures 12 and 13 plot radio flux density against values of the inner
disk radius.  Figure 12 plots inner disk radii derived from
phenomenological fits, while Figure 13 shows the inner disk radii
derived from fits including the relativistically-blurred reflection
model.  In these figures, it is clear that individual measures of the
inner disk radius generally lie within a very narrow range of values.
It is also clear that variations in the radio flux density occur
without changes in the inner disk radius.  This lack of a clear
correlation is confirmed in rank correlation tests against the line
and reflection--derived radii, and the disk normalization from which
continuum--based radii derive.  There is a single exception that
merits additional scrutiny:

In fits including the relativistically--blurred reflection model, the
inner disk normalization (radius) measured through continuum fits is
anti-correlated with radio flux density: $\rho = -0.533$,
corresponding to a 1.5\% chance of a false correlation.  In this
instance, the correlation is skewed by just two data points, derived
in fits to the spectra obtained on 02 and 04 June (see Table 4).
There is no evidence of a general trend in the normalization (radius)
values in question.  As noted previously, data taken at that time are
particularly affected by calibration problems in the 1.5--3.5 keV band
(for a discussion of similar problems, see Nowak et al.\ 2011).

Given that (1) only two points have driven the results of the rank
correlation test, (2) those two points are derived in spectra that are
less reliable than most other spectra, (3) fits to the disk continuum
with a more phenomenological model do not find a significant
correlation ($\rho = 0.312$, with a 18\% chance of false
correlation), (4) only four radius measurements are statistically inconsistent with a very
narrow range ($3 GM/c^{2} \leq r_{in} \leq 4 GM/c^{2}$), (5)
  systematic errors are likely to be much larger than statistical
  errors, and (6) independent fits to the Fe K line and reflection
  spectrum find no correlation with radio flux density, this apparent
  anti-correlation is not robust.

Taken as a whole, the data show that the jet power (as traced by radio
flux) is likely not modulated by changes in the inner radius of the
accretion disk in the low/hard state.

\subsection{Other Modes of {\it Disk}--Jet Coupling}
Although variations in the inner disk radius do not appear to be the
source of fluctuations in the jet, there is evidence of {\it
  disk}--jet coupling in our observations.  In the phenomenological
fits to the X-ray spectra, the disk flux and disk temperature are
positively correlated with the radio flux density ($\rho = 0.636$ and
$\rho = 0.615$, respectively, implying 0.3\% and 0.4\% probability of
false correlation).  In the more physical spectral fits, wherein a
relativistially--blurred disk reflection component is included, the
disk temperature and flux are the two parameters that are most
strongly correlated with the radio flux ($\rho = 0.675$ and $\rho =
0.640$, respectively, implying 0.1\% and 0.2\% probability of false
correlation).  Figure 14 plots radio flux density versus the disk
continuum flux measured in the phenomenological and more physical
models, while Figure 15 plots radio flux density versus the disk
temperature in each family of models.  It is notable that all of these
correlations are stronger than the correlation between radio flux and
the HXD/pin count rate, which is indicative of the correlations found
when low--resolution X-ray monitoring data or gas spectrometer data are used to
study disk--jet connections.

Simple accretion disk theory and fundamental work on
advection-dominated flows shows that the mass accretion rate may serve
to modulate the disk temperature, flux, and its inner radial extent
(e.g. Esin, McClintock, \& Narayan 1997).  Above a critical threshold
in mass accretion rate, however, it is likely that the disk is always
close to the innermost stable circular orbit.  This appears to be
largely confirmed by recent observations with CCD and dispersive X-ray
spectrometers, especially after disk correction factors are considered
(e.g., Reynolds \& Miller 2012).  The critical point seems to be
$L/L_{Edd} \simeq 0.001$, corresponding to $\dot{m}_{Edd.} \simeq
0.01$ for a canonical accretion efficiency of 10\% (Miller et
al.\ 2006, Rykoff et al.\ 2007, Tomsick et al.\ 2009, Reis, Fabian, \&
Miller 2010).

The luminosity of Cygnus X-1 is over this threshold: the mean
bolometric luminosity, based on the values in Table 3, is $L =
1.2\times 10^{37}~ {\rm erg/s}$, or $L/L_{\rm Edd} \simeq 0.006$.
Thus it is possible that variations in the mass accretion rate
primarily affect the disk temperature and flux rather than the inner
disk radius.  This is supported by the relative constancy of the disk
radii that we have derived in both phenomenological and more physical
spectral models, using both the disk continuum and the disk reflection
spectrum (see Figures 12 and 13).  The jet in the low/hard state may
be sensitive to the only disk parameters that {\it can} vary in this
regime.  It is particularly appealing that the disk flux -- a proxy
for the mass accretion rate through the disk -- is positively
correlated with radio flux.  This could indicate a simple coupling
between the mass accretion rate in the disk and the power of the jet.

It is also interesting to note that the disk ionization (as measured
by the reflection model) shows a weak positive correlation with the
radio flux density ($\rho = 0.415$, indicating a 6.9\% chance of false
correlation.  This parameter is difficult to measure, and errors are
typically large.  Nevertheless, a plot of radio flux density versus
the ionization parameter appears to show a weak positive trend (see
Figure 16).

%---------------------------------------------------

Again, prior correlations between radio flux density and
X-ray flux from monitoring data are effectively a correlation between
the radio flux density and the {\it hard} X-ray flux.  Those
correlations make no statement about the origin of the hard X-ray
flux, nor how the hard X-ray flux might be divided between different
components.  Table 5 shows that radio flux density is
correlated with the indices and flux of the (broken) power-law component
used in phenomenological spectral fits to the {\it Suzaku} data.  This
provides a relatively straightforward basis for the correlation
between radio flux density and hard X-ray flux (see Figures 5 and 17).

However, the broken and cut-off power-law model used in the
phenomenological fits is really a proxy for a full reflection model.
Correlations using the more physical model parameters give the
opposite result: radio flux density is anti-correlated with the {\it
  direct} power-law flux ($\rho = -0.414$, corresponding to a 7\%
chance of false correlation).  Instead, the radio flux density is
positively correlated with the {\it reflected} flux ($\rho = 0.551$,
giving a 1.2\% chance of false correlation; see Table 7 and Figure
18).  In this context, it is also useful to build on the results in
Section 4.3 by noting that the reflected flux is found to be {\it
  anti-correlated} with the direct power-law flux ($\rho = -0.662$,
giving a 0.2\% chance of false correlation; see Figure 19), in
apparent contradiction of simple disk reflection models.  As discussed
below, this finding is fully consistent with a model involving light
bending effects (e.g. Miniutti \& Fabian 2004).

\section{Discussion}
We observed Cygnus X-1 in a standard low/hard state on 20 occasions
with {\it Suzaku} during 2009, while making frequent monitoring
observations with the AMI radio telescope (see Tables 1 and 2).
Spectral fits to the X-ray data employed both phenomenological models,
and more physical disk reflection models (see Tables 3 and 4).
Correlations were then calculated between parameters derived in the X-ray
spectral fits, and the radio fluxes obtained via the monitoring
program (see Tables 5--8).  This program marks the first time that a
mission covering a broad X-ray bandpass, but also offering moderate
spectral resolution and low energy coverage via a CCD spectrometer,
has made monitoring observations capable of detecting direct and
reflected emission from the cool disk.  Numerous measures were taken
to ensure the best possible study of the disk, including explicit
modeling of the ionized companion wind, the use of an irradiated disk
model, and the use of a blurred reflection model suited to high
ionization.  A number of potential insights into disk--jet coupling,
the nature of the hard X-ray corona, and interactions between the
disk, corona, and jet are discussed below.

A central result of this program is that variations in power of the
relativistic jet (as traced by the radio flux) do not appear to be
driven by variations in the inner disk radius, at least not within the
confines of the sampling and methods employed in this program.  This
result is based on the absence of significant correlations between
inferred disk radii and radio flux (see Tables 3--6).  However, an
early indication of this result is already clear from a visual
inspection of Figure 4: the radio flux shows much higher fractional
variability than the X-ray flux.  Given that disk radii depend on the
square root of the disk flux normalization, the fractional variability
in the disk normalization would have to be {\it very} strong in order
to match the radio flux variations.  The spectral fits reveal only
small changes in the disk normalization, indicating only small changes
in the disk radius.  This result is corroborated via fits to the
relativistic iron disk line, and fits to the full disk reflection
spectrum, which are completely independent measures of the disk
radius.  The inner disk radii derived in those fits are also
uncorrelated with radio flux, and lie within narrow ranges (see Tables
3, 4, 5, and 7, and Figures 12 and 13).

The constancy of the inner disk in fits to the thermal continuum of
LMC X-3 in the high/soft state, for instance, was recently explained
in terms of the disk remaining close to the ISCO (Steiner et
al.\ 2010).  The relative stability of the inner disk radii derived in
our X-ray spectral fits (see Tables 3 and 4, and Figures 12 and 13) is
also an indication that the inner disk tends to sit near the innermost
stable circular orbit.  Indeed, the best fits strongly point to radii
consistent with the ISCO around a spinning black hole, i.e. within
$6~GM/c^{2}$.  In this regard, Cygnus X-1 is likely not unique.  A
growing number of results from spectrometers with low-energy coverage
suggest that disks may remain at the ISCO in the low/hard state, down
to $L/L_{\rm Edd} \simeq 0.001$ (Miller et al.\ 2006, Rykoff et
al.\ 2007, Reis et al.\ 2010; Reynolds et al.\ 2010, Reynold \& Miller
2011, Duro et al.\ 2011).  Observations also indicate that disks may
begin to radially recede as per the predictions of ADAF models at or
below this threshold (Tomsick et al.\ 2009).  In its low/hard state,
Cygnus X-1 is a factor of a few above this critical Eddington
fraction, consistent with fitting results suggesting that the disk is
close to the ISCO.

Given that a steady, compact jet operates in the low/hard state of
Cygnus X-1 while the disk appears to remain close to the ISCO, it is
possible that the truncation of a standard thin disk may not be required
in order to drive relativistic jets.  Truncation may aid in creating
stable poloidal magnetic fields, for instance (Reynolds, Garofalo, \&
Begelman 2006), and for such a reason may amplify jet power, but the
results of this program suggest that truncation is not a requirement.
Taken at face value, our results suggest that a standard thin
accretion disk that extends to the ISCO can drive relativistic jets.
This action may be mediated by the hard X-ray corona, or course, and
the onset of jet production as black holes transit from high/soft or
intermediate states into the low/hard state may have more to do with
changes in the corona than changes in the disk.

The second central result of this program is that the accretion disk
is connected to the relativistic jet: in both phenomenological and
physical spectral fits, the disk flux and disk temperature are found
to be correlated with radio flux.  These parameters are likely tied to
the mass accretion rate through the disk, and may serve as evidence of
a connection between the accretion rate in the disk and the radio
luminosity of the jet.  Jet radio luminosity is an imperfect proxy for
jet power (Allen et al.\ 2006), but this result may also nominally indicate a
link between the mass accretion rate in the disk and jet power.

It is also important to note and explore connections between the jet
and the hard X-ray corona.  Radio flux is found to correlate
positively with parameters such as the photon index of the power-law
spectral component.  Our results also suggest that the corona may have
an indirect a role, at least, in driving relativistic jets.  Like
other studies that require a disk close to the ISCO in the low/hard
state, our results suggest that the hard X-ray corona is not a
central, optically-thin, Comptonizing region {\it within} a truncated
disk.  This does {\it not} rule out Comptonization; indeed, the high
energy turn-over seen in these spectra is a hallmark of Comptonization
(see, e.g., Makishima et al.\ 2008 and Torii et al.\ 2011).  Rather, a
different geometry is merely required to describe these observations.

Beloborodov (1999) developed a model for accretion and ejection in
Cygnus X-1, wherein hard X-rays are generated in mildly relativistic
plasma moving away from the disk.  Especially in light of later
work by Markoff, Falcke, \& Fender (2001) and Markoff, Nowak, \& Wilms
(2005), it is natural to associate the outflowing relativistic plasma
with the base of the radio jet.  In the Beloborodov (1999) model, the
ejected material is initially lifted above the disk by magnetic
flaring activity generated by the magneto-rotational instability
(MRI), and may be ejected by the pressure of reflected radiation.
Thus, this model predicts a connection involving the disk, corona,
{\it disk reflection}, and the jet.

As noted previously, positive correlations are found between disk flux
and temperature, and radio flux.  These disk parameters are likely
tied to the mass accretion rate, which is of course linked to the MRI.
Thus, these simple correlations are consistent with one aspect of the
plasma ejection model of Beloborodov (1999).  However, a different
correlation may offer even deeper insights and more specific support:
the data show a positive correlation between the {\it reflected} X-ray
flux and the radio flux (see Figure 18).  A weaker correlation between
disk ionization and radio flux is also found; see Table 7 and Figure
16).  The positive correlation between reflected flux and radio flux
is consistent with the prediction that plasma blobs are ejected partly
due to the pressure of the {\it reflected} radiation.  Note that
reflected flux strongly dominates over direct disk flux in all of our
spectral fits (see Table 4).  Thus, our results may offer specific,
detailed support for the plasma ejection framework described in
Beloborodov (1999).

A further prediction of the plasma ejection model is that the X-ray
power-law spectral index, $\Gamma$, is related to the velocity of the
outflowing plasma via $\Gamma = 1.9/\sqrt B$, where $B =
\gamma(1+\beta)$ (here, $\beta = v/c$ and $\gamma = 1
/\sqrt{1-\beta^2}$).  Fits to the {\it Suzaku} data with blurred
reflection models are consistent with a power-law index of $\Gamma =
1.4$, corresponding to $v/c \simeq 0.3$.  It is worth noting that this
velocity is the escape speed for a distance of $z\simeq 20~GM/c^{2}$
from the black hole.  It is also notable that this velocity is broadly
consistent with the plasma velocity ($v/c = 0.4$) assumed at the base
of the jet in Markoff, Nowak, \& Wilms (2005).  In that work, emission
from the base of a jet is shown to naturally include Comptonization,
and shown to generate disk reflection that is consistent with
observations.  An important aspect of these broadband jet models is
that injected particles are accelerated into a power-law tail at a
height of $z=$10--100~$GM/c^{2}$ above the disk (see below).

Although fits with the phenomenological spectral models appear to
confirm prior evidence of feedback between the disk and corona in the
form of the $R-\Gamma$ correlation (Zdziarski et al.\ 2003; see
Figures 10 and 11, and Tables 5 and 6), more physical and
self-consistent fits with the blurred disk reflection spectrum find an
anti-correlation between the incident and reflected flux (see Section
4.3, Tables 6 and 7, and Figure 19).  This can be explained in terms of
gravitational light bending close to a spinning black hole.  Depending
on the vertical displacement of hard X-ray emission relative to a
black hole, more or less hard flux may be beamed onto
the disk, leading to a non-linear relationship between direct and
reflected flux (Miniutti \& Fabian 2004).  This sort of non-linear
relationship has been observed in some Seyfert-1 AGNs, including
MCG-6-30-15 (Minutti et al.\ 2007), but also in the stellar-mass black hole XTE
J1650$-$500 (Miniutti, Fabian, \& Miller 2004; Rossi et al.\ 2005).  

Relativistic blurring and light bending may also explain one point of
disparity between the results of this program and Beloborodov (1999).
The outflowing corona model was partly motivated by results suggesting
low reflection fractions in stellar-mass black holes in the low/hard
state.  The model was developed before extremely blurred lines were
observed in stellar-mass black holes using CCD spectrometers, however,
and before potential evidence of light bending had been observed in
flux trends.  After blurring, a reflection model is not as sharp, and
a higher reflection flux may be required to fit peaks in the data.
Moreover, if light bending is even somewhat important, it is difficult
to ascertain how much flux is seen directly versus how much is focused
onto the disk.  Although most of our fits are not consistent with a
low reflection fraction, the basics of outflowing corona models do not
depend on the reflection fraction.

The results of fits to the disk continuum, disk line, and reflection
spectrum are given in Section 4.  Factoring in all of the models and
results, and emphasizing fits where strong constraints are obtained,
the data suggest that Cygnus X-1 has a spin in the range of $0.6 \leq
a \leq 0.99$.  This rather large range derives primarily from lower
spin values found based on fits to the disk continuum.  However, the
models used were less physical than the relativistic reflection models
that were employed.  The reflection fits strongly prefer a
near-maximal spin for Cygnus X-1.  This is at odds with the results of
recent reflection fits to some {\it XMM-Newton} spectra of Cygnus X-1
(Miller et al.\ 2009); however, those spectra may give falsely low
spin values owing to photon pile-up effects (see, e.g., Miller et
al.\ 2010).  However, the high spin implied in our fits is consistent
with more recent and more detailed work by Gou et al.\ (2011), which
places a $3\sigma$ lower limit of $a > 0.91$.  It is also consistent
with recent work by Duro et al.\ (2011), which measures $a =
0.88^{+0.07}_{-0.11}$ using a new mode of the {\it XMM-Newton}/EPIC-pn
camera, and a measurement of $a = 0.97^{+0.01}_{-0.02}$ based on fits
to {\it Suzaku} spectra (Fabian et al.\ 2012).  Thus, anisotropic hard
X-ray emission due to gravitational light bending, if not also bulk
flows (see above), is plausible in Cygnus X-1.

In this work, an emissivity of $r^{-3}$ was held fixed in all fits
with ``Laor'' and ``kdblur''.  This action was taken partly because
steeper indices were not required by the data when considering a
simple power-law functional form, and partly because spin
estimates made by Duro et al.\ (2011) also suggest this emissivity is
reasonable.  However, an index of $q=3$ is at odds with the steeper
indices sometimes obtained in black holes where fits suggest high spin
values, such as MCG-6-30-15 (Brenneman \& Reynolds 2006; Miniutti et
al.\ 2007,), 1H 0707-495 (Fabian et al.\ 2009, Wilkins \& Fabian
2011), XTE J1650-500 (Miller et al.\ 2009), GX 339$-$4 (Miller et
al.\ 2008, Reis et al.\ 2008), and XTE J1550$-$564 (e.g. Miller et
al.\ 2009; Steiner et al.\ 2011).  It is interesting to note that
emissivity may be positively correlated with Eddington fraction, since
most of the high $q$ values are recorded in ``intermediate'' and
``very high'' states (for a review of the stellar-mass black hole
cases, see Miller et al.\ 2009).

% A potential difficulty with this interpretation is that fits with the
% Laor disk line model, and blurred reflection model, were all broadly
% consistent with a standard disk emissivity of $q = 3$ (where $J(r)
% \propto r^{-q}$; see Table 3 and 4).  The spin estimates made by Duro
% et al.\ (2011) also suggest this emissivity is reasonable.  Indeed,
% this consistency led to a value of $q=3$ being fixed in all spectral
% fits.  This particular value is at odds with the higher indices often
% obtained in black holes where fits suggest high spin values, such as
% MCG-6-30-15 (Brenneman \& Reynolds 2006; Miniutti et al.\ 2007,), 1H
% 0707-495 (Fabian et al.\ 2009, Wilkins \& Fabian 2011), XTE J1650-500
% (Miller et al.\ 2009), GX 339$-$4 (Miller et al.\ 2008, Reis et
% al.\ 2008), and XTE J1550$-$564 (e.g. Miller et al.\ 2009; Steiner et
% al.\ 2011).  In the case of the stellar-mass black holes, it is
% interesting to note that emissivity may be positively correlated with
% Eddington fraction, since most of the high $q$ values are recorded in
% ``intermediate'' and ``very high'' states (for a review of the
% stellar-mass black hole cases, see Miller et al.\ 2009).

Motivated by the disk reflection spectrum seen in 1H 0707$-$479
(Fabian et al.\ 2009), Wilkins \& Fabian (2011) have calculated how
the emissivity index should vary with black hole spin for different
idealizations of the hard X-ray corona.  It is found that the
emissivity can be represented as a twice-broken power-law, with a
break radius at or within $6~GM/c^{2}$.  Especially if the black hole
spins rapidly and the hard X-ray source is nearby, the inner
emissivity index is expected to be steep ($q>3$), while the outer
index is expected to be flatter.  However, if the source of hard X-ray
emission is $z \simeq$~10--30$~GM/c^{2}$ above a spinning black hole,
$q=3$ is expected.  This model implies that the source of hard X-ray
emission in the low/hard state of Cygnus X-1 may be approximately
10--30~$GM/c^{2}$ above the disk.  Preliminary fits with the broken
power-law emissivity index suggest that Cygnus X-1 may follow the
predictions of Wilkins \& Fabian (2011); a fit to a single spectrum of
Cygnus X-1 assuming an on-axis point source is detailed in Fabian et
al.\ (2012).  Our assumption of a less extreme emissivity is
consistent with a corona that has a small but finite spatial extent.

At a height of $\simeq 20~GM/c^{2}$, the light bending model
independently predicts that reflected flux should be anti-correlated
with the incident power-law flux (see the top panel of Figure 2 in
Miniutti \& Fabian 2004).  Figure 19 plots the reflected flux measured
in our spectral fits, versus the measured power-law flux.  The
anti-correlation between these parameters recorded in Table 8 is clear
in the plot as well.  At heights greater than $\simeq 20~GM/c^{2}$,
light bending is predicted to be inefficient (Miniutt \& Fabian 2004).
Thus, if light bending plays a role in driving the variability and
reflection seen in Cygnus X-1, a height of $z \simeq 20~GM/c^{2}$ is
required; both larger and smaller values are inconsistent with the
observed flux pattern.  It is not clear how else the anti-correlation
between reflected and incident flux could be explained; changes in the
disk ionization are insufficient and would produce a positive
correlation as the disk albedo increases with ionization.

Thus, taken as a whole, the results of this program suggest a simple
and self-consistent picture that can be summarized in stages: (1)
Material in the corona (jet base) may be initially lifted from the
disk by MRI.  This is consistent with observed correlations between
radio flux and disk flux and temperature.  (2) The material can then
be accelerated at least partially through the pressure of reflected
radiation.  This is supported by a correlation between radio flux and
reflected X-ray flux.  (3) Comptonization in the plasma above the disk
produces the hard X-ray spectrum in a manner that links the bulk
velocity of the outflow and the X-ray power-law index.  The observed
power-law indices are suggestive of a moderately relativistic plasma,
with $v/c \simeq 0.3$.  (4) Last, the relationship between direct and
reflected hard X-ray flux, and the disk emissivity index, are driven
by the distance between the hard X-ray source and black hole.  The
reflection fitting results bear the hallmarks of light bending and
suggest hard X-ray emission from $z\simeq 20~GM/c^{2}$ above the black hole.
This distance coincides with the zone where hard X-rays are generated
in jet-based models for black hole spectral energy distributions
(Markoff, Nowak, \& Wilms 2005), and is consistent with the lag
timescales uncovered in advanced X-ray variability studies of Cygnus
X-1 (Uttley et al.\ 2011; also see Wilkinson \& Uttley 2009).

The results discussed here imply that hard X-ray emission is produced
in a fairly compact region.  It is difficult to make an independent
test of the size of the emission region in stellar-mass black holes;
however, microlensing makes such tests possible in AGN.  Recent
microlensing results in quasars find that the hard X-ray emission
regions have half-light radii of only 10~$GM/c^{2}$ (Chartas et
al.\ 2009, Dai et al.\ 2010; Chen et al.\ 2011).  These results
provide some important observational support for compact, central,
hard X-ray emission regions.

Our results offer some insights into the nature of the disk,
the hard X-ray corona, and jet production in stellar-mass black holes
in the low/hard state.  Moreover, they serve to highlight some natural
connections between the plasma ejection model of Beloborodov (1999),
broadband jet models (e.g. Markoff, Nowak, \& Wilms 2005), and the
effects of gravitational light bending on hard X-rays produced close
to a spinning black hole (e.g. Miniutti \& Fabian 2004).  Can this
confluence of models and processes also explain the absence of jets in
other black hole states?

As noted above, high values of the disk emissivity index are
preferentially found in the ``intermediate'' and ``very high''
spectral states.  In these phases, disk reflection is especially
strong.  Both of these properties are consistent with hard X-ray
emission originating closer to the black hole, perhaps at $z
\simeq 5-10~ GM/c^{2}$, where the escape velocity is higher, making it
harder to eject a plasma.  Moreover, the Beloborodov (1999) coronal
ejection model predicts a lower plasma velocity for the steeper
spectral slopes that are typically observed in ``intermediate'' and
``very high'' spectral states.  Reflection is weakest, and jets are
weakest (or quenched entirely) in disk--dominated ``high/soft'' states
(see, e.g., Russell et al.\ 2011), where power-law fits to hard X-ray
spectra generally yield the steepest slopes.  This is again consistent
with clues to the nature of disk--corona--jet coupling that emerge
from Cygnus X-1.

Livio, Pringle, \& King (2003) describe a model for disk--jet coupling
in stellar-mass black holes and AGN that relies on fluctuations in the
poloidal magnetic field emerging from the accretion disk.  This model
builds upon the work of Blandford \& Payne (1982) but makes specific
predictions and comparisons to individual sources such as GRS
1915$+$105.  Our results are broadly compatible with this model,
especially if it is a combination of the disk flux ($\dot{m}$),
temperature, and ionization that serve to modulate poloidal magnetic
fields.  This model does not address the role of disk reflection in
setting disk ionization, nor in helping to accelerate gas that has
been lifted off of the disk.  Thus it is not necessarily in conflict
with the models described by Beloborodov (1999), Markoff, Nowak, \&
Wilms (2005), and Miniutti \& Fabian (2004).  It is possible that
radiation pressure from reflected radiation and magnetocentrifugal
forces may work in tandem to drive jet plasma away from the disk.

In many respects, Seyfert-1 AGN resemble stellar-mass black holes in
the low/hard or ``very high'' state: their power-law indices are
broadly similar, disk reflection spectra are prominent in many cases,
and power spectra show band--limited noise with characteristic
frequencies.  In terms of their Eddington fraction, however, most
Seyfert-1 AGN would be better associated with the ``low/hard'' state
(see, e.g., Blustin et al.\ 2005).  It is interesting to briefly
consider how the results of this study of Cygnus X-1 might impact our
understanding of Seyfert AGN.

Unlike stellar-mass black holes, where the flux in hard X-rays or
reflected X-rays can dominate the total luminosity, UV emission from
the much cooler disk around a supermassive black holes dominates the
radiative output of Seyferts.  Radiation pressure from the disk itself
should strongly dominate over pressure from reflected emission.  At
present, no comparable study of a Seyfert has been performed that
could confirm or reject the correlation between reflected X-ray flux
and radio flux found in Cygnus X-1.  However, coordinated studies of
AGN from different classes are now being undertaken.  Recent efforts
to obtain contemporaneous X-ray and radio monitoring observations of
the Seyfert-1 NGC 4051, for instance, reveal a mode of disk--jet
coupling that may differ from stellar-mass black holes in the low/hard
state (King et al.\ 2011; Jones et al.\ 2011).  Note, however, that
jet-based spectral models do provide a good description of the
broadband SED of NGC 4051 (Maitra et al.\ 2011).  Low-luminosity AGN
(LLAGN) may be a better comparison to Cygnus X-1 in the low/hard
state, although it is more difficult to study the inner accretion disk
in such sources.  Nevertheless, radio and X-ray flux monitoring of the
LLAGN NGC 7213, for instance, has shown a coupling similar to that
revealed in stellar-mass black holes in the low/hard state (Bell et
al.\ 2011).

Last, it is worth remarking on the limited scope of this analysis, and
offering some brief comparisons to other recent efforts to model black
hole spectra.  Nowak et al.\ (2011) have shown that a variety of
simple phenomenological models, complex Comptonization models, and
disk reflection models can all provide good fits to spectra of Cygnus
X-1.  Both disk reflection and Comptonization must be at work, at some
level, but models differ as to the degree and the specifics of the
accretion flow geometry.  This analysis has focused on reflection; it
is not perfectly consistent in its treatment of Comptonization and the
roll-over at high energy.  Indeed, fully consistent descriptions of
Comptonization prove to be difficult.  Some recent efforts to model
spectra from Cygnus X-1 have taken the extraordinary step of using two
Comptonization components: the electron temperatures in two ``compps''
components are linked, but each region has a distinct optical depth
(e.g. Makishima et al.\ 2008).  If the gravitational potential sets
the local electron temperature, then these regions must be close to
each other.  Their different optical depths imply different densities
and internal pressures, however, and they would diffuse into each
other on a short interval driven by the local orbital timescale.  The
regions could potentially be kept distinct via some form of magnetic
confinement, but that evokes the geometry that is implied by our fits
-- one of a corona with a strong magnetic component.  Though different
families of models still differ considerably, it is possible that they
are on a convergent path.

\section{Conclusions}

$\bullet$ Variations in the radial extent of the inner accretion disk
in the low/hard state of Cygnus X-1 do not appear to drive changes in
the relativistic jet.

$\bullet$ However, the flux and temperature of the inner accretion disk --
parameters likely tied to $\dot{m}$ -- positively correlate with radio
flux, suggesting a true {\it disk}--jet coupling.

$\bullet$ The stability of the inner disk in the low/hard state in
Cygnus X-1 suggests that it sits close to the ISCO.  This is supported
by the detailed results of direct spectral fits.

$\bullet$ The implication of a disk close to the ISCO in the low/hard
state would signal that standard thin disks {\it can} power
relativistic jets.

$\bullet$ Factoring in both phenomenological and more physical fits to
the disk continuum and reflection spectrum, and making cuts on
measurement quality, the inner disk radii suggest a spin of $0.6 \leq
a \leq 0.99$, consistent with some other recent results (Gou et
al.\ 2011, Duro et al.\ 2011; also see Miller et al.\ 2005).  Blurred, ionized
reflection fits prefer a near-maximal spin.

$\bullet$ Some fundamental predictions of a simple disk reflection
geometry appear to be confirmed in phenomenological X-ray spectral
fits.  More physical and more self-consistent fits with a blurred
reflection spectrum may require an anisotropic hard X-ray source,
among other complexities.

$\bullet$ A positive correlation between the {\it reflected} X-ray
flux and the radio flux suggests a connection between disk reflection
and jet power.  This is consistent with a plasma ejection model for
the corona (or, jet base) developed by Beloborodov (1999).

$\bullet$ An anti-correlation between the reflected and direct
power-law flux is consistent with a model in which gravitational light
bending close to a black hole partially drives the flux variations
(e.g. Miniutti \& Fabian 2004).  The anti-correlation between these
fluxes suggests a source height of $z \simeq 20~GM/c^{2}$, which is
consistent with an independent prediction based on the disk emissivity
parameter.

$\bullet$ Hard X-ray production at a height of $z \simeq 20~GM/c^{2}$
is consistent with recent time lags found in X-ray variability studies
(Uttley et al.\ 2010), and consistent with theoretical predictions for
the region of initial particle acceleration in the base of a jet
(Markoff, Nowak, \& Wilms 2005).

%\section{Acknowledgements}

\hspace{0.1in}

We thank the anonymous referee for comments that have improved this
manuscript.  We thank Koji Mukai, Robert Petre, Kazuhisa Mitsuda, and
the entire Suzaku team for their help in executing these observations.
We thank Elena Gallo, Dipankar Maitra, Chris Reynolds, Mark Reynolds,
and Mateusz Ruszkowski for helpful discussions.

\clearpage

\begin{table}[t]
\caption{Radio Observations and Flux Densities}
\begin{footnotesize}
\begin{center}
\begin{tabular}{lllll}
MJD & ${\rm F}_{15}$~(mJy) & ~ &  MJD & ${\rm F}_{15}$~(mJy) \\
% & MJD & Flux Density & MJD & Flux Density \\
\tableline

 54895.4000244 &  11.2(2) & ~ & 55080.7900391 &  12.6(2)  \\
 54899.3200073 &   7.2(1) & ~ & 55081.8100586 &  17.2(2)  \\
 54901.5599976 &  14.0(2) & ~ & 55085.0500488 &  10.6(1)    \\
 54903.2999878 &  11.1(1) & ~ & 55086.8699951 &  15.4(1) \\
 54906.2299805 &   8.7(2) & ~ & 55089.6800537 &  11.4(4) \\
 54907.5499878 &  13.2(5) & ~ & 55090.8699951 &  12.0(2) \\
 54910.2399902 &  11.4(1) & ~ & 55092.7700195 &  11.9(2) \\
 54912.2500000 &   8.9(2) & ~ & 55095.9399414 &   8.7(1) \\
 54913.1900024 &  13.8(2) & ~ & 55099.7500000 &  12.8(2)  \\
 54917.2899780 &  13.2(1) & ~ & 55101.7299805 &  12.1(1)  \\
 54919.3099976 &  15.0(1) & ~ & 55105.6999512 &  15.3(1)  \\
 54920.2299805 &  14.3(2) & ~ & 55106.6700439 &  10.1(2) \\
 54923.3400269 &  14.1(2) & ~ & 55107.8499756 &   9.3(2) \\
 54928.3200073 &   9.3(2) & ~ & 55108.8000488 &  11.2(1) \\
 54933.1900024 &  11.6(1) & ~ & 55109.9300537 &  13.3(2) \\
 54934.2899780 &  13.6(2) & ~ & 55113.8000488 &   8.7(2) \\
 54940.1400146 &   8.8(1) & ~ & 55114.6700439 &  12.8(2) \\
 54945.1699829 &  10.8(1) & ~ & 55117.6800537 &  10.2(1) \\
 54947.1500244 &  12.3(2) & ~ & 55119.6800537 &  13.7(1) \\
 54948.2199707 &  12.3(1) & ~ & 55121.8800049 &  16.3(2) \\
 54953.4000244 &  10.2(2) & ~ & 55122.9300537 &  10.0(1) \\
 54954.4000244 &  10.6(2) & ~ & 55123.9200439 &   9.1(1) \\
 54956.4199829 &  12.5(4) & ~ & 55124.5300293 &   8.6(7) \\
 54958.0599976 &   9.6(1) & ~ & 55125.6199951 &   9.4(8) \\
 54960.3800049 &  11.0(2) & ~ & 55129.9000244 &   5.9(2) \\
 54961.3800049 &   8.8(2) & ~ & 55130.9000244 &   6.5(1) \\
 54962.4000244 &  12.2(2) & ~ & 55131.7900391 &   6.4(2) \\
 54963.3900146 &  17.3(1) & ~ & 55132.9000244 &   9.1(1) \\
 54971.1199951 &  13.5(1) & ~ & 55133.8900146 &  10.5(2) \\
 54972.4000244 &  15.7(7) & ~ & 55136.8499756 &   6.8(1) \\
 54976.2899780 &  13.1(3) & ~ & 55137.8800049 &  10.6(2) \\
 54983.3200073 &  14.7(2) & ~ & 55138.5100098 &   10(2) \\
 54987.1300049 &  19.6(1) & ~ & 55141.5100098 &   7.2(3) \\
 54989.3099976 &  23.0(6) & ~ & 55143.7199707 &   9.3(1) \\
 54994.2000122 &  15.3(3) & ~ & 55145.8299561 &   5.1(8) \\
 54998.2199707 &  26.7(3) & ~ & 55146.8399658 &   6.1(2) \\
 54999.2000122 &  29.3(2) & ~ & 55147.8699951 &   5(1) \\
 55000.1500244 &  16.7(2) & ~ & 55148.4799805 &   8(2) \\
 55001.2600098 &  17.9(2) & ~ & 55150.8800049 &  10.1(1) \\
 55002.2600098 &  23.1(2) & ~ &  55152.6400146 &   6.0(1) \\
 55004.2800293 &  25.8(3) & ~ &  55155.5600586 &  10.9(6)\\
 55005.2500000 &  24.2(3) & ~ &  55158.5200195 &  11.6(3)\\
 55008.0700073 &  23.9(4) & ~ &   55161.4499512 &  13.3(2)  \\
 55011.1199951 &  21.5(2) & ~ &   55162.4899902 &  12.8(2) \\
 55011.9500122 &  17.9(2) & ~ &  55163.7299805 &  18.1(1)  \\
 55012.8800049 &  13.5(3) & ~ &  55166.6600342 &  13.6(1)  \\
 55014.1199951 &  22.0(2) & ~ &  55168.6899414 &   7(1) \\
 55015.2299805 &  20.4(3) & ~ &  55176.4699707 &  16.2(2)  \\
 55016.1199951 &  15.3(1) & ~ &  55178.7600098 &  12.8(1)  \\
 55020.1699829 &  16.8(1) & ~ &  55187.7399902 &  15.3(5)  \\
 55035.1099854 &  11.5(1) & ~ &  54965.1300049 &  13.9(1)  \\
 55043.9699707 &  14.9(1) & ~ &  55003.0999756 &  26.6(1)  \\
 55046.0699463 &  12.7(2) & ~ &  55017.0999756 &  13.7(1)  \\
 55047.0300293 &  16.4(4) & ~ &  55063.7199707 &  14.3(2)  \\
 55050.0600586 &  12(1)   & ~ &  55085.9100342 &  13.6(1)  \\
 55052.0400391 &  18.3(8) & ~ &  55176.5899658 &  14.5(1)  \\
 55055.9100342 &  13.6(2) & ~ &  55196.4599609 &  14.4(2)  \\
 55058.9100342 &  14.0(1) & ~ &  55194.4699707 &  14.0(5)  \\
 55063.0400391 &  13.7(2) & ~ &  55193.4599609 &  16.7(1)  \\
 55072.8800049 &   6.6(1) & ~ &  55192.6700439 &  12.5(1)  \\
 55076.6899414 &   7(1)   & ~ &  55190.6099854 &  12.0(2)  \\
 55078.9200439 &  11.1(1) & ~ &  55189.4799805 &  16.9(3)  \\
\tableline
\end{tabular}
\vspace*{\baselineskip}~\\ \end{center} 
\tablecomments{High resolution imaging of Cygnus X-1 shows that it has
  a compact jet in its low/hard state (Stirling et al.\ 2001).  In our
  2009 campaign, the activity of this jet was traced through its
  15~GHz flux density, as measured in frequent monitoring observations
  made with the AMI.  The start times and flux densities measured in
  AMI observations that spanned our {\it Suzaku} campaign our given
  above.}
\vspace{-1.0\baselineskip}
\end{footnotesize}
\end{table}

\clearpage

\begin{table}[t]
\caption{X-ray Observations}
\begin{footnotesize}
\begin{center}
\begin{tabular}{llll}
MJD     &  Month, Day &   XIS used  &    Exposure (ks) \\
\tableline

54924.0537384 & Apr 03 & 0$+$3 & 15.5  \\
54929.2562847 & Apr 08 & 0$+$3 & 13.1 \\
54935.7650463 & Apr 14 & 0$+$3 & 13.0 \\
54944.1674769 & Apr 23 & 1     & 16.3 \\
54949.7099884 & Apr 28 & 0$+$3 & 12.4 \\
54957.7006829 & May 06 & 0$+$3 & 15.3 \\
54971.0242593 & May 20 & 0$+$3 & 17.5 \\
54976.3582060 & May 25 & 0$+$3 & 16.1 \\
54980.4956713 & May 29 & 0$+$3 & 25.7 \\
54984.4814005 & Jun 02 & 0$+$3 & 15.0 \\
54986.8208333 & Jun 04 & 0$+$3 & 6.8 \\
55125.3772107 & Oct 21 & 0$+$3 & 15.1 \\
55130.2652315 & Oct 26 & 0$+$3 & 20.0 \\
55138.8946181 & Nov 03 & 0$+$3 & 14.1 \\
55145.8193982 & Nov 10 & 0$+$3 & 18.2 \\
55152.2855093 & Nov 17 & 0$+$3 & 18.7 \\
55159.5142245 & Nov 24 & 0$+$3 & 14.0 \\
55166.2922454 & Dec 01 & 0$+$3 & 19.8 \\
55173.6473495 & Dec 08 & 0     & 17.3 \\
55182.0619329 & Dec 17 & 0     & 3.4 \\

\tableline
\end{tabular}
\vspace*{\baselineskip}~\\ \end{center} 
\tablecomments{The table above lists the start time of our 2009 {\it
    Suzaku} observations of Cygnus X-1 in UT, as well as the calendar day on which the
  observation was made, which XIS units were used in spectral fits,
  and the total exposure time.  The start times and exposure times are
  drawn from the HXD/pin event files as multiple XIS event files exist
  owing to frequent switching between editing modes.}
\vspace{-1.0\baselineskip}
\end{footnotesize}
\end{table}

%\begin{table}
%\caption{Phenomenological X-ray Spectral Model Parameters}
%\begin{footnotesize} 
%\begin{center}
%\begin{tabular}{llllllllllllllllllll}

\begin{turnpage}
\begin{deluxetable}{llllllllllllllllllll}
\tablefontsize{\footnotesize} 
%\tabletypesize{\scriptsize}
%\rotate \tablewidth{0pt}
%\tablecaption{Phenomenological X-ray Continuum Spectral Parameters}
%\tablefontsize{\footnotesize}
\tablehead{
\colhead{Date}	& \colhead{$\chi^{2}$} &  \colhead{${\rm N}_{\rm H}/10^{21}$}  &  \colhead{kT}  &  \colhead{K$_{\rm disk}$}  & \colhead{${\rm R}_{\rm disk}$} &  \colhead{$\Gamma_1$} &  \colhead{$\Gamma_2$}  &  \colhead{${\rm E}_{\rm break}$} &  \colhead{K$_{\rm PL}$}  &  \colhead{${\rm E}_{\rm cut}$} &  \colhead{${\rm E}_{\rm fold}$}  &    \colhead{${\rm E}_{\rm Laor}$} &  \colhead{${\rm R}_{\rm line}$}  &  \colhead{K$_{\rm Laor}$}  &  \colhead{EW}  &   \colhead{0.8-10.0} &  \colhead{0.8-500}  & \colhead{disk} \cr   
\colhead{} & \colhead{} & \colhead{($cm^{-2}$)} & \colhead{(keV)} & \colhead{$(10^{5})$} & \colhead{${\rm GM}/{\rm c}^{2}$} & \colhead{} & \colhead{} & \colhead{(keV)} & \colhead{} & \colhead{(keV)} & \colhead{(keV)} & \colhead{(keV)} & \colhead{${\rm GM}/{\rm c}^{2}$} & \colhead{($10^{-2}$)} & \colhead{(eV)} & \colhead{($10^{-8}$)} & \colhead{($10^{-8}$)} & \colhead{($10^{-8}$)} } 
\startdata
\hline
Apr03 & 2430.2  &	1.3(3)	&    0.198(3) &   1.5(1) &  3.3(2) &	1.76(1)	 & 1.40(1) & 6.4(1) &	2.33(3) & 22(1) &  153(3) &  6.70$^{+0.01}$ & 4.6(6) &  1.5(2) & 176.0  &  1.40  &    3.38   &    0.135 \\

Apr08 &  2282.8 	& 1.0$^{+0.1}$ &  0.195(3) &   1.35(7) &	 3.1(2) &  1.776(5) & 1.40(1) & 7.1(1) &	1.90(1) & 24(1) &  163(4) & 	6.90(4) &       3.7(4) & 1.3(2) & 210.0 &   1.11  &    2.62  &     0.107 \\

Apr14 &  2455.0	 & 1.3(3)	 &   0.191(2) &	1.2(1) &  2.9(3) &	1.763(7) & 1.41(1) & 7.2(2) &	1.58(1) & 24(1) &  160(4)  &   6.89(3) &       3.6(4) & 1.1(1) & 217.0 &   0.93 &     2.20   &    0.082 \\

Apr23 &	2701.8 &	12(2) &	0.180(5) &	0.8(1) &  2.4(4) &	1.683(8) & 1.41(1) & 7.3(5) &	1.53(1) & 23(1) &  159(3)  & 	6.96(2) &  4(1) &   1.1(2) & 187.0 &   0.94  &    2.37  &     0.035 \\
					
Apr28 &  2397.9	 & 5(1) &	0.203(4) & 1.1(2) &  2.8(5) &	1.74(1)	& 1.43(1) & 7.9(1) &	2.36(3) & 24(2) &  167(4) &  6.94(3) &       3.4(3) & 1.4(2) & 150.0 &   1.12  &    2.53  &     0.137 \\

May06 &  2492.4	 & 1.0$^{+0.1}$ &  0.203(2) &	1.60(6) &  3.3(1) &   1.853(3)  & 1.42(1) & 6.8(1) & 2.33(1) &  23(1) &  152(3) & 	6.92(3)	  &     3.7(2) &  1.6(1) & 252.0 &   1.31  &    2.89 &	0.165 \\
			
May20 &  2369.2	 & 2.0(1)	&     0.204(2) &	1.80(8)	&  3.5(2) &  1.892(8) & 1.47(1) & 6.6(1) & 2.29(2) & 22(1) &  169(4) &  6.89(3)  &      3.3(3) &  1.5(1) & 264.0 &   1.26 &      2.63 &      0.185 \\
			
May25 &	2507.5  & 3.4(8) &	 0.209(2) &	2.30(7)	& 4.0(1) & 1.965(7)  & 1.45(1) &  6.5(1) & 3.20(2) & 21(1) &  136(3) &  6.89(3)  &      3.4(2) &  2.2(2) & 300.0 &   1.69 &     3.41  &     0.288 \\

May29 &  3075.9	 & 1.0$^{+0.1}$ &  0.210(1) &	3.49(7)	& 4.9(1) & 2.121(2) & 1.52(1) & 6.0(1) & 3.95(1) &  23(1) &   129(2) &  6.8(1) & 3.4(1) &  2.4(1) &  340.0  &  1.92  &    3.43	&      0.442 \\

Jun02 &	2617.4  &	2(1) &	0.217(3) &	2.7(2) & 4.3(3) &	2.10(1)	 & 1.56(1) & 6.4(1) & 4.00(4) & 22(1) &   136(3) &   	6.88(5) &	       3.3(3) &  2.2(2) &  318.0 &   1.95 &     3.40 & 	 0.426 \\

Jun04 &	2659.8  &	2.6(6) &	0.214(3) &	2.6(1)  &  4.3(2) 	& 2.135(5) & 1.60(1) & 5.1(1) &	3.89(3) & 31(1) &  165(5) &    6.73(3) &   3.0(3) &  2.3(2) & 308.0 &   1.82   &   3.29 &	 0.368 \\
 
Oct21 &  2373.6	 & 1.1(1) &	0.199(2) &	1.48(8)	&  3.2(2) &  1.839(5) & 1.44(1) & 6.6(1) & 1.74(1) & 22(1) &  160(4) &  6.8(1) &  4.0(3) &   1.0(1) & 196.0 &   1.00 &     2.20 &	      0.130 \\
			
Oct26 &	2429.8 	& 1.4(4) &	0.193(3) &	1.29(7)	& 3.0(2) &  1.802(5) & 1.42(1) &  6.8(1) & 1.67(1) & 22(1) &  172(4) &  6.86(4)  &      3.8(3) &  1.0(1) &  206.0 &   0.96  &    2.21  &     0.093 \\

Nov03 &	1969.4  &	6(2) &	0.180(7) &	1.1(3) &  2.7(8) &	1.67(1) & 1.41(1) &	8.6(8) & 1.16(2) & 23(1) &  172(6) &  	6.8(1)	&       3(1) & 0.5(1) & 100.0  &  0.72   &   1.78 &	 0.046 \\

Nov10 &	2559.7  &	2(1) &	0.132(6) &	3.8(7)	& 5.2(9) &   1.620(4) & 1.41(1)  &  9.3(5)  &	0.95(2) & 24(1) &  170(4) & 	6.73(3)	 &      1.6(3) & 0.4(1) & 86.0 &   0.61  &    1.57 &	 0.015 \\

Nov17 &	2407.6  &	3(1) &	0.208(3) &	1.00(9)	& 2.6(2) &   1.782(6) & 1.39(1) & 7.0(1) & 2.49(2) & 24(1) &  153(3) & 	6.92(4)	 &      3.8(4) & 1.7(1) & 214.0  &  1.44  &    3.43 &	 0.124 \\

Nov24 &	2301.9  &	1.5(5) &	0.206(2) &	1.35(9)	& 3.1(2) &  1.84(1) & 1.45(1) & 6.5(2) &	2.09(2) & 23(1) &  175(3) & 	6.87(4)	 &       3.4(3) &  1.4(1) & 231.0  &  1.19  &    2.18 &	 0.156 \\
			
Dec01 &	2335.9  &	1.1(1) &	0.200(2) &       1.3(4)	& 3.0(9) &   1.795(7) & 1.42(1) & 6.7(2) &	 1.88(2) & 23(1) & 166(4)  &  6.93(3)  &      4.4(2)  & 1.1(1) &  192.0 &   1.11  &    2.54  &     0.128 \\

Dec08 &	2680.6  &	10(3) &	0.192(7) &	0.9(3)	& 2.5(6) &   1.70(1) 	 & 1.41(1) & 8.7(8) & 1.75(2) & 23(1) & 170(4) &  	6.94(3)	 &      3.4(6) & 1.1(2)  & 182.0  &  1.06   &   2.56	&      0.058 \\
			
Dec17 &	2566.1  &	5(2) &	0.220(4) &	1.36(9)	& 3.1(2) &   1.94(1)	 & 1.43(1) & 5.9(1) & 3.94(4) & 24(1) & 153(5) &   6.90(5)  &      3.1(2) & 3.3(2) &  331.0  &  2.07   &   4.48 &      0.274 \\

\hline
\enddata
\tablecomments{The table above lists the values of variable parameters
  in phenomenological fits to the {\it Suzaku} spectra of Cygnus X-1.
  After binning the XIS to a minimum of 10 counts per bin, each of the
  above combined spectra (XIS, HXD/pin, HXD/GSO) had 2421 degrees of
  freedom.  The total phenomenological spectral model used was:
  const$\times$tbabs$\times$mtable\{wind\_abs\}$\times$
  (gauss+laor+diskpbb+bknpow)$\times$highecut (see Section 3.1).  The
  third column is the equivalent Hydrogen column density of the {\it
    ionized} companion wind within the binary system.  The fourth and
  fifth columns give the temperature and flux normalization of the
  disk continuum component; the sixth column gives an estimate of the
  inner radius based on the assumptions and correction factors
  described in Section 3.1.  Columns 7--12 list the values measured
  from the broken and cut-off power-law continuum component (column 10
  gives the flux normalization factor).  Columns 13--16 give values of
  the Laor relativistic line function used to describe the broad Fe K
  emission line detected in the spectra.  Columns 17--19 give the
  unabsorbed flux measured in the soft X-ray band (0.8--10.0~keV), full
  fitting band (0.8--500.0~keV), and total disk continuum flux, in
  units of $10^{-8}~ {\rm erg}~ {\rm cm}^{-2}~ {\rm s}^{-1}$.  Owing
  to space limitations, flux errors are not given; however, the error
  in the flux is directly proportional to the error in the continuum
  component normalizations.}
\end{deluxetable}
%\tablecomments{Each of the above fits have 2421 degrees of freedom.}
\end{turnpage}
%\end{tabular}
%\vspace*{\baselineskip}~\\ 
%\end{center} 
%\tablecomments{Comments go here.}
%\vspace{-1.0\baselineskip}

\begin{turnpage}
\begin{deluxetable}{llllllllllllllllllll}
\tablefontsize{\footnotesize} 
%\rotate \tablewidth{0pt}
%\tabletypesize{\scriptsize}
\tablecaption{Relativistic X-ray Disk Reflection Spectral Parameters}
%\tablefontsize{\footnotesize}
\tablehead{
\colhead{Date}	& \colhead{$\chi^{2}$}  &  \colhead{$\nu$}  &  \colhead{${\rm R}_{\rm refl}$} & \colhead{$\xi$} & \colhead{${\rm K}_{\rm refl.}$} & \colhead{kT} & \colhead{${\rm K}_{\rm disk}$} &  \colhead{${\rm R}_{\rm disk}$} & \colhead{$\Gamma$} & \colhead{${\rm K}_{PL}$} & \colhead{${\rm E}_{\rm cut}$} & \colhead{${\rm E}_{\rm fold}$} & \colhead{0.8--500.0} & \colhead{disk} & \colhead{PL} & \colhead{refl.} \cr   
\colhead{} & \colhead{} & \colhead{} & \colhead{${\rm GM}/{\rm c}^{2}$} & \colhead{($10^{3}$)} & \colhead{($10^{-5}$)} & \colhead{(keV)} & \colhead{($10^{5}$)} & \colhead{${\rm GM}/{\rm c}^{2}$} & \colhead{~~~~} & \colhead{~~~~} & \colhead{(keV)} & \colhead{(keV)} & \colhead{($10^{-8}$)} & \colhead{($10^{-8}$)} & \colhead{($10^{-8}$)} & \colhead{($10^{-8}$)}}

\startdata
\hline
Apr03 & 2729.5 &        2424 &	1.2(2) &	3.0(3) &	1.6(1) & 	0.185(2) &	2.0(1) &  3.7(2) &	1.40(1) &   0.58(2) &	   100(3) &	  182(9) &	6.24	 &       0.11   &  4.06 &	2.07 \\

Apr08 & 2484.1 &	2424 &	1.6(3) &	3.0(2) &	1.5(1) &	0.177(2) &	2.1(2) &  3.8(2) &	1.40(1)  &     0.40(1)  &     102(4) &    210(10) &   	 4.88 &	       0.08  &   3.00 &	1.80 \\

Apr14 & 2592.0 &	2424 &	1.5(3) &	3.0(4) &	1.2(1) &	0.172(4) &	2.1(2) &  3.8(2) &	1.40(6) &      0.36(2)  &      100(4) &    195(10)  &   4.17 &	       0.06  &   2.65 &	1.45  \\	

Apr23 &	2794.8 &	2424 &	1.8(3) &	1.6(2) &	1.5(2) &	0.145(7) &	2.2(5) &  3.9(3) &	1.49(1) &      0.79(3)   &     95(4) &    190(10) &   	 4.78 &	       0.02  &   3.91 &	0.86  \\
					
Apr28 & 2527.4 &	2424 &	1.6(2) &	3.3(4) &	1.2(2) &	0.177(3) &	1.6(2) &  3.3(4) &	1.40$^{+0.02}$  &     0.42(3)  &      96(4)  &   215(20)  &      4.69   &        0.06  &   2.98  &    1.65  \\

May06 & 2675.5 &	2424 &	20(9) &	        3.8(4) &	1.3(1) &	0.183(2) &	2.7(2) &  4.3(1) &	1.47(7) &     0.30(3) & 147(10) &	   236(30) & 	 4.45 &	       0.14  &   1.81  &	2.55  \\
			
May20 & 2500.3 &	2424 &	9(6) &	        5.0(4) &	1.1(1) &	0.190(2) &	2.3(2) &  4.0(1) &	1.40$^{+0.07}$  &    0.15(1)   &    170(10)  &   244(40)  &     4.00    &        0.17   &   1.26    &   2.57  \\
			
May25 &	2069.9 &	1878 &	8(2) &	        3.8(3) &	1.8(1) &	0.220(4) &	1.9(1) &  3.6(1) &	1.400$^{+0.03}$  &    0.10(1)  &     162(10)  &   174(40)  &    4.85    &        0.29   &   1.16    &   3.40  \\

May29 & 2444.5 &	1878 &	1.6(2) &	4.4(3) &	1.7(1) &	0.237(3) &	1.8(2) &  3.5(1) &	1.44(2) &	0.155(5) &	   174(10) &	   94(20) &   4.60  &	       0.57   &   0.96  &	3.07  \\

Jun02 &	2012.0 &	1878 &	1.4(1) &	4.0(3) &	1.8(1) &	0.240(2) &	1.4(1) &  3.1(4) &	1.41(1)  &    0.20(2)  &    155(10)  &    169(30)  &  	 4.69	  &       0.56   &   1.14  &	3.00  \\

Jun04 &	2149.7 &	1878 &	1.4(2) &	4.4(4) &	1.6(1) &	0.230(2) &	1.3(1) &  3.0(2) &	1.41(1)  &  0.17(2)  &    157(20)  &   165(40)  &   	 4.69	  &       0.49   &   1.24  &	3.00  \\
 
Oct21 & 2537.8 &	2424 &	1.2(2) &	4.2(3) &	0.81(7) &	0.182(2) &	2.2(2) &  3.9(1) &	1.41(1) &    0.19(2) &	  148(10) &	   185(30)  &    3.44	  &       0.11    &  1.42  &	1.90  \\
			
Oct26 &	2653.0 &	2424 &	4(2) &	        3.8(4) &	1.0(1) &	0.169(3) &	2.3(2) &  4.0(1) &	1.41(1)  &   0.27(2)   &   130(5)   &    197(10)  &     3.87     &       0.07    &  2.05   &    1.72  \\

Nov03 &	1981.9 &	2424 &	1.8(5) &	2.2(3) &	1.4(3) &	0.14(3) &	2.3(2) &  4.0(6) &	1.45(2) &	0.34(5)	 &   109(9)  &    240(30)  &   	 3.29	   &      0.02    &  2.10  &	1.16  \\

Nov10 &	2831.7 &	2424 &	1.5(1) &	3.8(8) &	0.30(5) &	0.1321)  &	4.5(6) & 5.6(9) &      1.48(1) &	0.57(2)	 &   93.0(6) &	   205(20) &	 3.45  &	  0.03   &   2.99  &	0.44  \\

Nov17 &	2618.8 &	2424 &	6(3) &	        2.3(2) &	2.7(1) &	0.17(1)	 &      3.0(5) &  4.6(1) &	1.44(1) &	0.40(2)	 &  146(7) &	   163(20) &	 5.44  &	   0.04    &  2.61  &	2.75  \\

Nov24 &	2407.2 &        2424 &	5(2) &	        3.9(2) &	1.2(1) &	0.190(2) &	1.8(2) & 3.5(2) &	1.41(1) &	0.22(2)	 &  153(7) &	   180(20) &	 4.17	  &       0.13    &  1.77   &	2.28  \\
			
Dec01 &	2316.5 &	2147 &	13(6) &	        3.1(3) &	1.2(1) &	0.179(2) &	2.6(2) & 4.3(7) &	1.46(1)  &     0.35(2)   &    146(7)   &   165(20)  &     4.10    &        0.10   &   2.15    &   1.85  \\

Dec08 &	2762.6 &	2424 &	2.1(3) &	3.3(3) &	1.1(2) &	0.14(1)  &	3(1) &	4.8(3) &        1.42(1) &	0.50(2)	 &  109(8) &	    192(20) &	   5.00	     &    0.03   &   3.37  &	1.60	  \\
			
Dec17 &	2332.0 &	2147 &	12(6) &	        3.5(1) &	2.4(1) &	0.211(3) &	1.8(9) & 3.5(2) &	1.40$^{+0.01}$   &    0.27(2)   &    154(20)  &    228(70)  &   6.68    &        0.25    & 2.23    &   4.19    \\

\hline
\enddata
\tablecomments{The table above lists the values of variable parameters
  in more physical relativistic disk reflection fits to the {\it
    Suzaku} spectra of Cygnus X-1.  The total spectral model employed
  in these fits is given by:
  const$\times$tbabs$\times$mtable\{wind\_abs\}$\times$
  (gauss+dispkbb+pow+kdblur$\times$atable\{reflionx.mod\})
  $\times$highecut (see Section 3.2).  Columns 4--6 give the
  parameters of the blurred reflection component, including the inner
  disk radius derived from the fits.  Columns 7--9 list disk
  parameters derived from fits to the disk continuum, including
  estimates of the radius based on our assumptions and corrections
  (see Sections 3.1 and 3.2).  Columns 10--13 give the parameters of
  the cut-off power-law continuum component.  Columns 14--17 list the
  total unabsorbed flux in the full fitting band (0.8--500.0~keV), as
  well as the flux in the disk continuum, power-law continuum, and
  reflection components, in units of $10^{-8}~ {\rm erg}~ {\rm
    cm}^{-2}~ {\rm s}^{-1}$.  Owing
  to space limitations, flux errors are not given; however, the error
  in the flux is directly proportional to the error in the continuum
  component normalizations.}
\end{deluxetable}
\end{turnpage}

\pagebreak

\centerline{~\psfig{file=f1.ps,width=5.0in,angle=-90}~}
\figcaption[h]{\footnotesize The figure above shows a deep {\it
    Chandra}/HETG spectrum of Cygnus X-1 in the low/hard state in 2003.  The
  narrow emission line at 6.40~keV is fitted with a simple Gaussian,
  but the Fe XXV and Fe XXVI absorption lines at 6.70 and 6.97~keV
  (respectively) are fitted with an XSTAR model with ${\rm N} =
  4.5\times 10^{21}~ {\rm cm}^{-2}$ and ${\rm log}(\xi) = 2.0$.  This
  ionization parameter permits a good fit in the Fe K band of this
  observation and is commensurate with that required by many ions in a
  comprehensive study of wind absorption with orbital phase (Hanke et
  al.\ 2009).  The same grid of ionized absorption models was applied to all {\it Suzaku}
  spectra of Cygnus X-1 treated in this paper.}
\medskip

\centerline{~\psfig{file=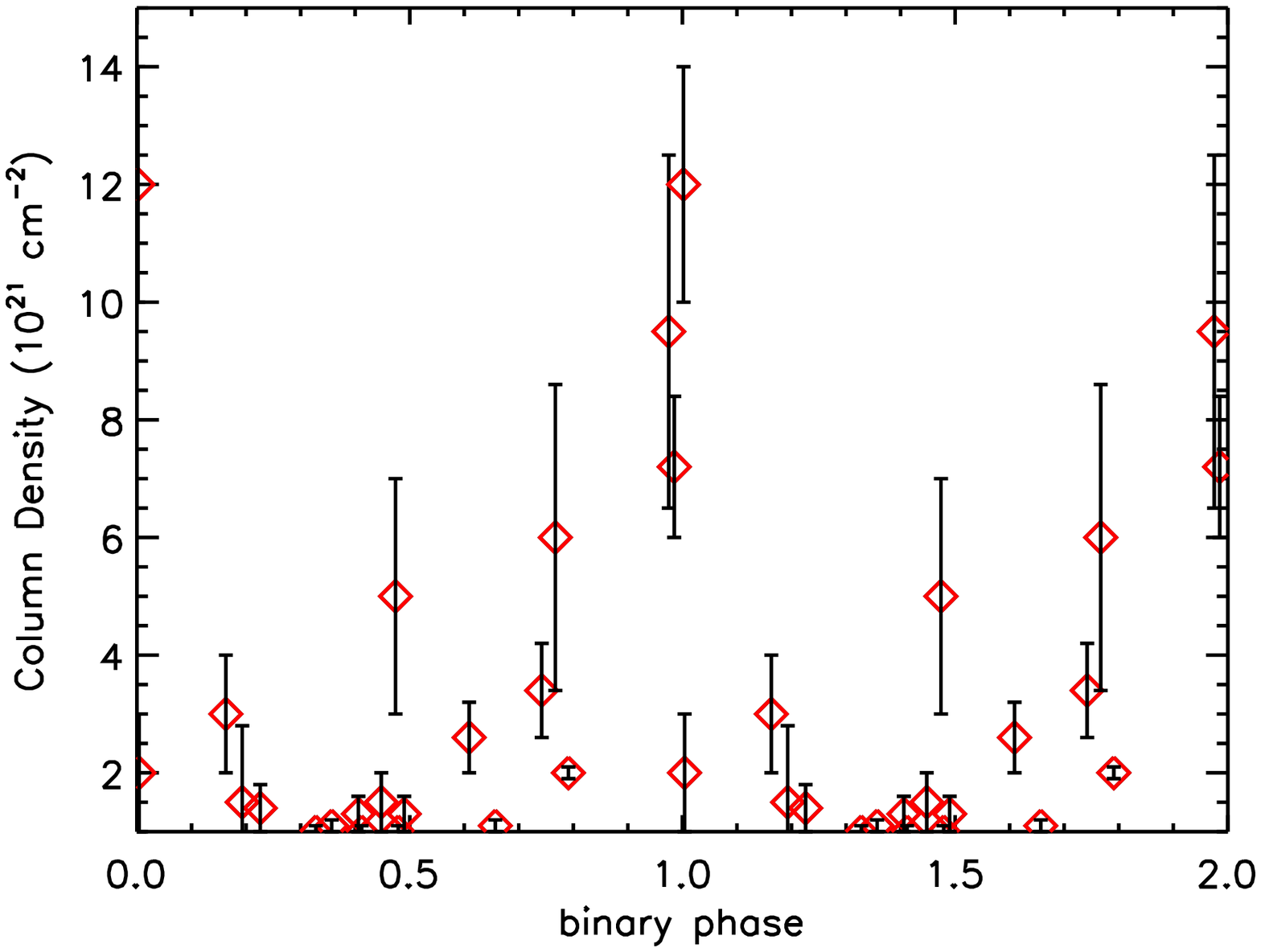,width=5.0in}~}
\figcaption[h]{\footnotesize To account for ionized absorption in the
  companion wind of Cygnus X-1, we constructed a grid of XSTAR models
  and included the grid as a table model with variable column density
  in all spectral fits.  This step was important to ensure that
  absorption did not falsely affect the disk continua nor the Fe K
  band.  The plot above shows that the ionized column varies as
  expected with phase: the absorption is generally strongest close to
  phase zero, when the O star companion is closest to the observer.}
\medskip

\clearpage

\centerline{~\psfig{file=f3.ps,width=5.0in,angle=-90}~}
\figcaption[h]{\footnotesize The panel above shows the combined
  backside-illuminated XIS (black) and HXD PIN (red) and GSO (blue)
  spectra of Cygnus X-1, obtained during an observation on 2009 May
  06.  The spectra have been fitted with a simple power-law, ignoring
  the range below 3~keV, the Fe K band (5--7~keV), and the 15--45~keV
  band.  When applied to the complete spectra extending down to
  0.8~keV, a cool accretion disk continuum and disk reflection
  features are clearly evident.  The spectra were binned for visual
  clarity.}
\medskip

\clearpage

\centerline{~\psfig{file=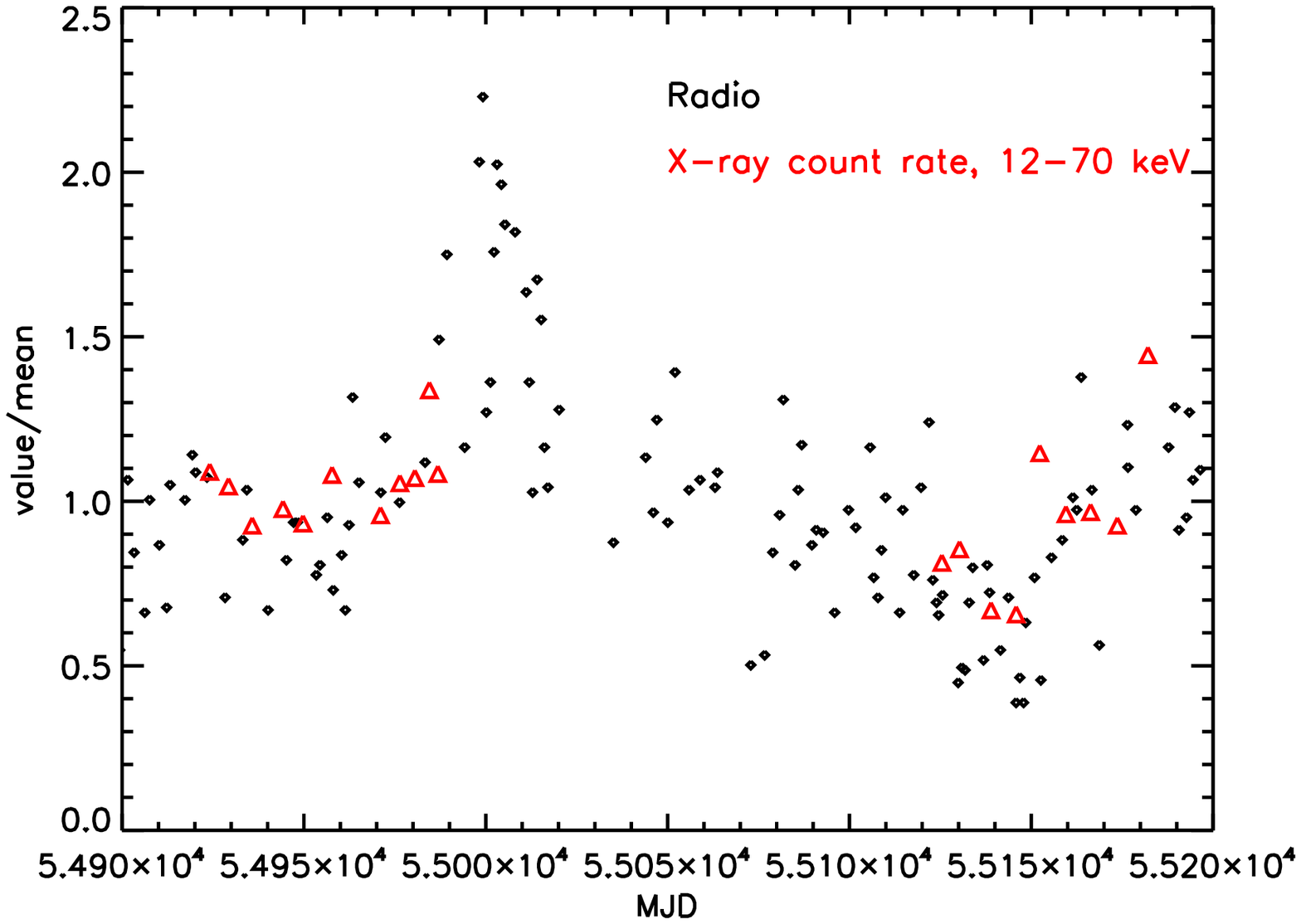,width=5.0in}~}
\figcaption[h]{\footnotesize The plot above shows time variations in
  the radio flux density and HXD/PIN count rate, relative to their
  mean values.  Prior studies of disk--jet connections have
  essenetially studied correlations between radio flux density and
  hard X-ray flux rather than disk properties, and these quantities
  are clearly correlated in our {\it Suzaku} observations.}
\medskip

\centerline{~\psfig{file=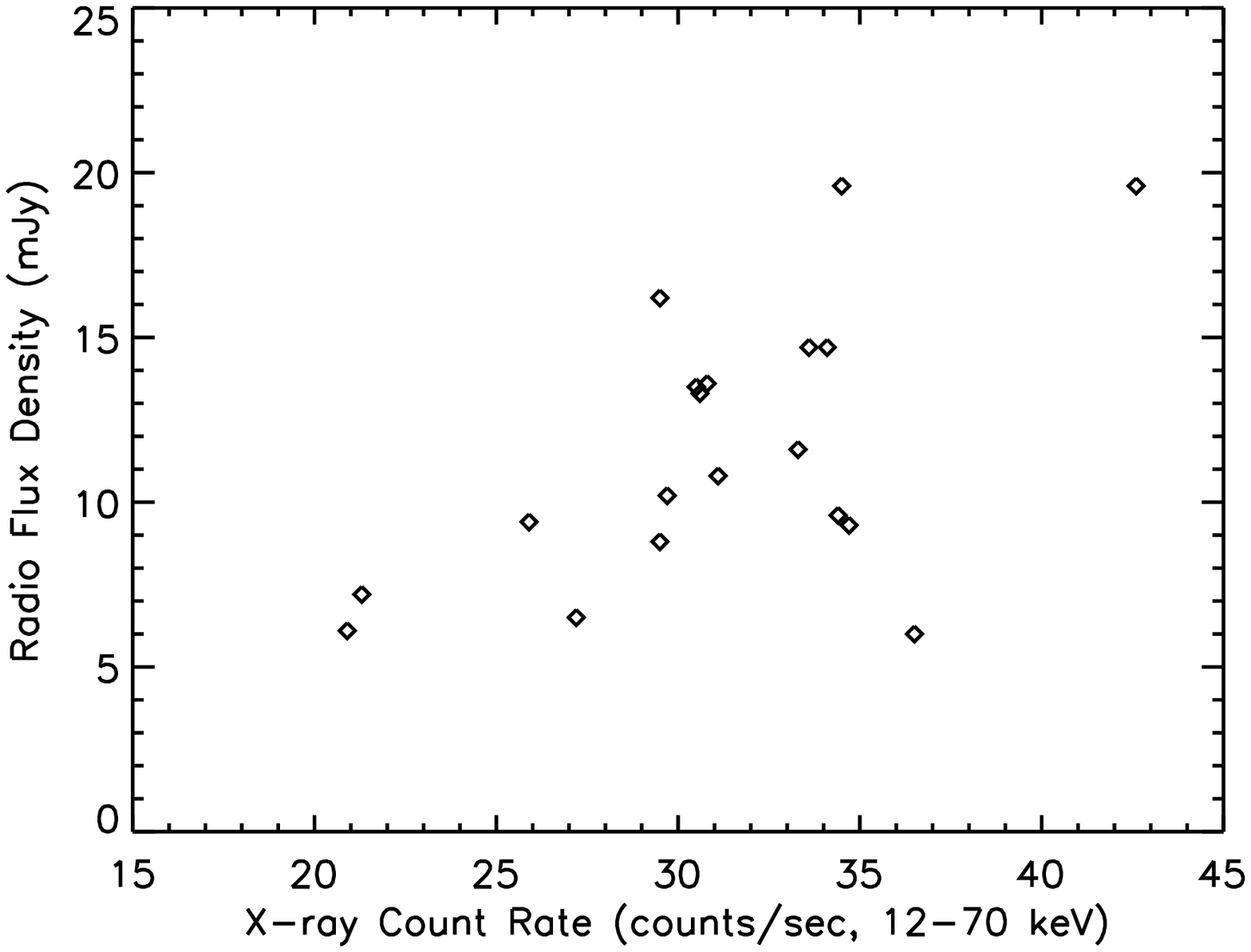,width=5.0in}~}
\figcaption[h]{\footnotesize This figure plots radio flux density
  versus X-ray count rate, for the 20 AMI observations most
  immediately following the 20 X-ray observations we made with {\it
    Suzaku}.  Errors on the radio flux density and X-ray count rate
  are similar to the size of the plotting symbols used in the figure.
  These model-independent quantities are positively correlated despite
  the obvious scatter: a Spearman's rank correlation gives $\rho =
  0.455$, and false correlation probability of 0.044.}
\medskip

\clearpage

\centerline{~\psfig{file=f6.ps,width=5.0in,angle=-90}~}
\figcaption[h]{\footnotesize The plot above shows spectra obtained on
  2009 April 14, fit with a phenomenological model.  Warm ionization was
  modeled using an XSTAR grid, the disk continuum was modeled using
  the simple ``diskpbb'' model, and the hard emission was modeled using
  a broken and cut-off power law.  The hard continuum approximates a
  reflection spectrum.  The Fe K band was modeled using a narrow
  Gaussian emission line at 6.40 keV, and a relativistic Laor emission
  line; Fe XXV and XXVI absorption lines from the XSTAR grid also
  affected the Fe K band.  The overall fit is good and all of the
  parameters are well-constrained.}
\medskip

\centerline{~\psfig{file=f7.ps,width=5.0in,angle=-90}~}
\figcaption[h]{\footnotesize The plot above shows the data/model ratio
  obtained when the flux of line components affecting the Fe K band is
  set to zero in the best-fit phenomenological model to spectra
  obtaind on 2009 April 14.  The narrow neutral Fe K line at 6.4~keV is
  clearly evident; ionized absorption in the Fe K  band is not very important in
  this particular spectrum.}
\medskip

\centerline{~\psfig{file=f8.ps,width=5.0in,angle=-90}~}
\figcaption[h]{\footnotesize The plot above shows the data/model ratio
  obtained when the Laor relativistic line component flux is set to
  zero in the best-fit phenomenological model to spectra obtaind on
  2009 April 14.  Narrow emission and absorption components have been
  fitted and only the relativistic component to emission in the Fe K
  band is shown.  The spectrum was binned for visual clarity.}
\medskip

\centerline{~\psfig{file=f9.ps,width=5.0in,angle=-90}~}
\figcaption[h]{\footnotesize The plot above shows spectra obtained on
  2009 Apr 14, fit with a physical blurred disk reflection model.
  Both dynamical and scattering broadening were included among the
  affects that shape the broad Fe K line.  Warm ionization was modeled
  using an XSTAR grid, the disk continuum was modeled using the simple
  ``diskpbb'' model.  The overall fit is good and all of the parameters
  are well-constrained.}
\medskip

\clearpage

\centerline{~\psfig{file=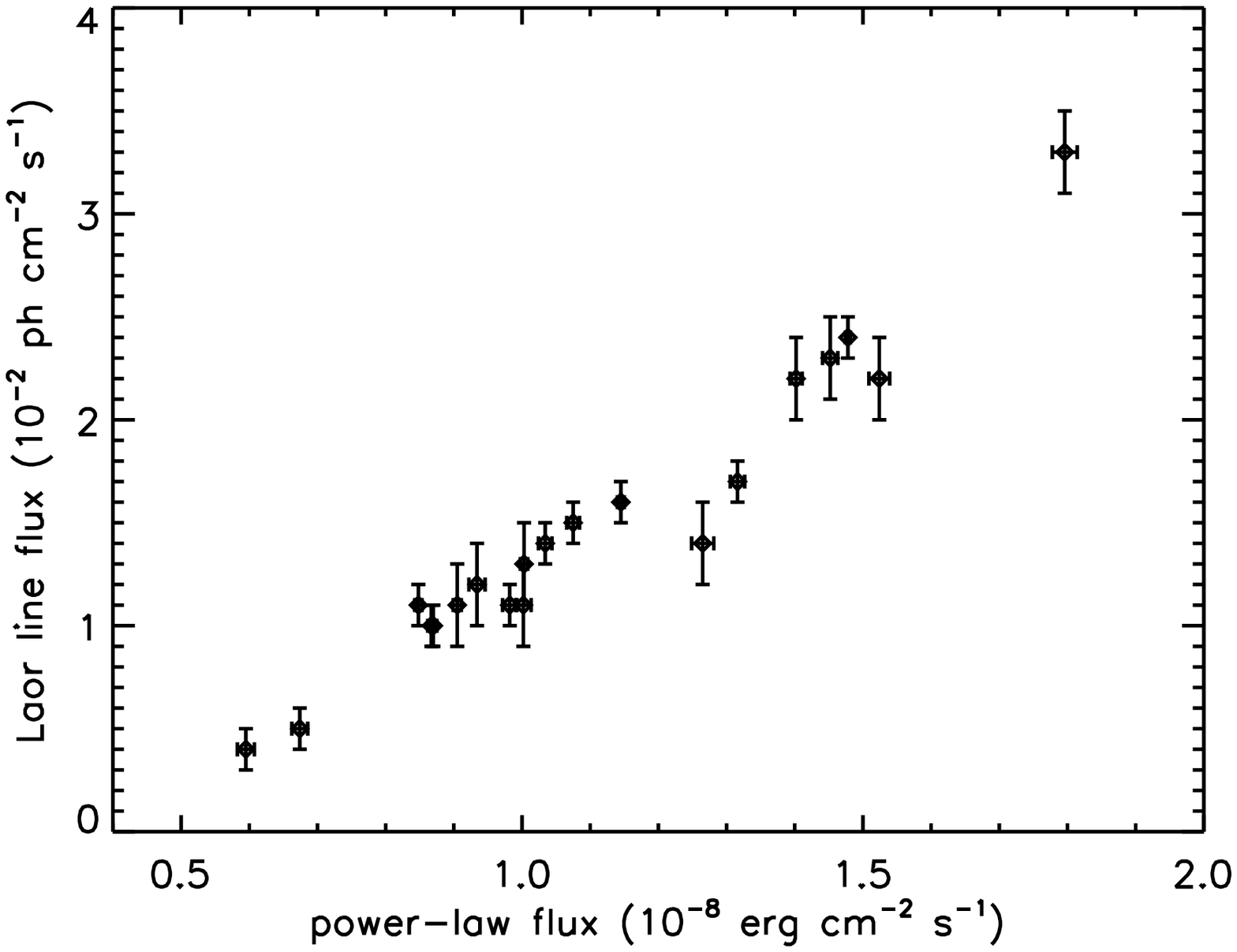,width=5.0in}~}
\figcaption[h]{\footnotesize If disk reflection is at work close to
  black holes, irradiation of the disk by external hard X-ray emission
  should stimulate an emission line.  The flux of the disk line should
  therefore be driven by, and closely linked to, flux represented by a
  power-law component.  The figure above plots relativistic disk line
  flux versus power-law flux for each of the 20 {\it Suzaku}
  observations of Cyg X-1.  The Spearman's rank correlation for these
  two parameters is $\rho = 0.958$, and the probability of a false
  correlation is $3.3\times 10^{-11}$.  The close link between the
  flux of the relativistic line and hard continuum strongly confirms
  that reflection from a disk extending close to the black hole is at
  work in the low/hard state of Cygnus X-1.}
\medskip

\centerline{~\psfig{file=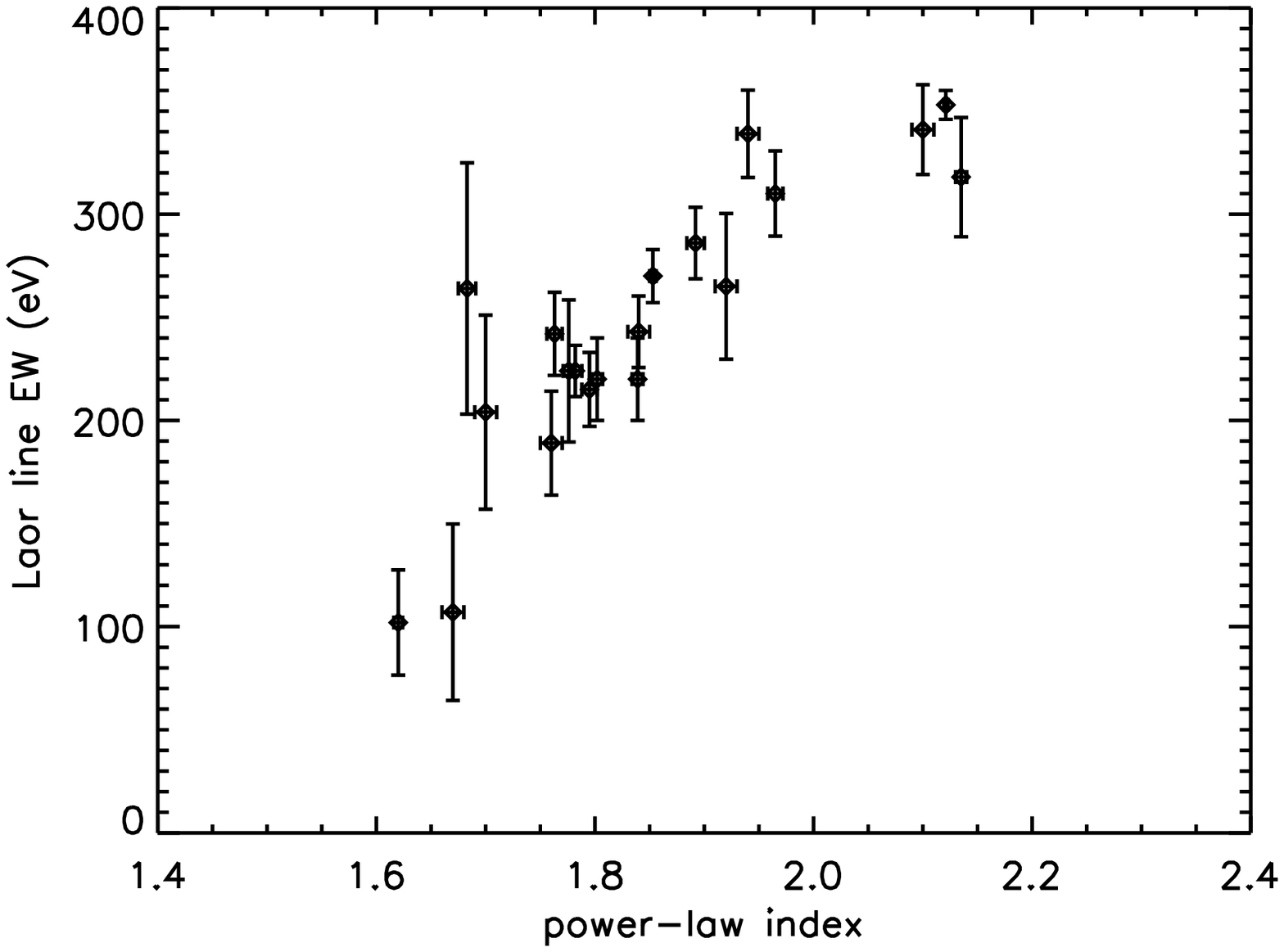,width=5.0in}~}
\figcaption[h]{\footnotesize In prior reflection studies, a
  correlation has been observed between the reflection fraction
  ($\Omega/2\pi$) versus the illuminating power-law index; this is
  taken as evidence of feedback between the corona and disk (Zdziarksi
  et al.\ 2003).  Since the equivalent width of a disk line and the
  reflection fraction are related for a given ionization (e.g. George
  \& Fabian 1991), the line equivalent width can be used as a proxy
  for the reflection fraction.  The plot above shows the relativistic
  Laor line equivalent width versus the soft X-ray power-law index
  ($\Gamma_{1}$) measured in our phenomenological fits.  As expected,
  the two are tightly correlated.  A Spearman's rank correlation test
  returns a coefficient of $\rho = 0.928$, and the probability of a
  false correlation is $3.9\times 10^{-9}$.  }
\medskip

\clearpage

\begin{table}[t]
\caption{Rank Correlations of Radio Flux and Phenomenological X-ray Model Parameters}
\begin{footnotesize}
\begin{center}
\begin{tabular}{llll}
Parameter 1 & Parameter 2 & Spearman's $\rho$ & False Alarm Probability	\\
\tableline
radio flux density & $\Gamma_{2}$ & 0.647 & 0.002 \\

radio flux density & power-law flux (0.8-10.0~keV) & 0.646 & 0.002 \\

radio flux density & disk flux & 0.636 & 0.003 \\

radio flux density & $\Gamma_{1}$ & 0.633 & 0.003 \\

radio flux density & disk kT &	0.615	& 0.004 \\

radio flux density & 0.8--10~keV flux &	0.610 & 0.004 \\

radio flux density & Laor flux & 0.606 & 0.005 \\

radio flux density & power-law norm. & 0.604 & 0.005 \\

radio flux density & Laor EW & 0.598 & 0.005 \\

radio flux density & ${\rm E}_{\rm break}$ & -0.555 & 0.011 \\

radio flux density & 0.8--500~keV flux & 0.491 & 0.028 \\

radio flux density & pin count rate & 0.455 & 0.044 \\

radio flux density & ${\rm R}_{in}$ (Laor) & -0.369 & 0.109 \\

radio flux density & disk norm.	& 0.312	& 0.180 \\

radio flux density & ${\rm E}_{\rm fold}$ & -0.263 & 0.262 \\

radio flux density & Laor Line Centroid & 0.151 & 0.524 \\

radio flux density & ${\rm E}_{\rm cut}$ & -0.012 & 0.958 \\

\tableline
\end{tabular}
\vspace*{\baselineskip}~\\ \end{center} 
\tablecomments{The table above lists the Spearman's rank correlation
  coefficient, $\rho$, and the associated false correlation
  probability, for different phenomenological X-ray fitting parameters
  versus radio flux density.}
\vspace{-1.0\baselineskip}
\end{footnotesize}
\end{table}

\vspace{1.0in}

\begin{table}[htb!]
\caption{Rank Correlations of Selected Phenomenological X-ray Model Parameters}
\begin{footnotesize}
\begin{center}
\begin{tabular}{llll}
Parameter 1 & Parameter 2 & Spearman's $\rho$ & False Alarm Probability	\\
\tableline
power-law flux & Laor flux & 0.958 & $3.3\times 10^{-11}$ \\

Laor equiv. width & $\Gamma_{1}$ & 0.928 & $3.9\times 10^{-9}$ \\

power-law flux & disk flux & 0.872 & $5.4\times 10^{-7}$ \\

disk flux & Laor flux & 0.865 & $8.5\times 10^{-7}$ \\

Laor equiv. width & $\Gamma_{2}$ & 0.729 & $2.7\times 10^{-4}$ \\

disk flux & disk norm. & 0.655 & 0.002 \\

disk norm. & disk kT & 0.452 & 0.045 \\

power-law norm. & disk norm. & 0.447 &	0.048 \\

power-law flux & ${\rm R}_{in}$ (Laor) & -0.175 & 0.462 \\

disk flux & ${\rm R}_{in}$ (Laor) & -0.147 & 0.537 \\

\tableline
\end{tabular}
\vspace*{\baselineskip}~\\ \end{center} 
\tablecomments{The table above lists the Spearman's rank correlation
  coefficient, $\rho$, and the associated false correlation
  probability, for different phenomenological X-ray fitting parameters.}
\vspace{-1.0\baselineskip}
\end{footnotesize}
\end{table}

\clearpage

\begin{table}[t]
\caption{Rank Correlations of Radio Flux and X-ray Reflection Model Parameters}
\begin{footnotesize}
\begin{center}
\begin{tabular}{llll}
Parameter 1 & Parameter 2 & Spearman's $\rho$ & False Alarm Probability	\\
\tableline

radio flux density & disk kT & 0.675 & 0.001 \\

radio flux density & disk flux & 0.640 & 0.002 \\

radio flux density & ${\rm E}_{\rm cut}$ & 0.600 & 0.005 \\

radio flux density & reflection flux & 0.551 & 0.012 \\

radio flux density & disk norm. & -0.533 & 0.015 \\

radio flux density & power-law norm. & -0.436 & 0.054 \\

radio flux density & $\xi$ & 0.415 & 0.069 \\

radio flux density & power-law flux & -0.414 & 0.070 \\

radio flux density & 0.8--500.0 keV flux & 0.357 & 0.122 \\

radio flux density & reflection norm. & 0.332 & 0.153 \\

radio flux density & ${\rm E}_{\rm fold}$ & -0.274 & 0.243 \\

radio flux density & $\Gamma$ & -0.204 & 0.388 \\

radio flux density & $R_{in}$ (reflection) & 0.090 & 0.707 \\

\tableline
\end{tabular}
\vspace*{\baselineskip}~\\ \end{center} 
\tablecomments{The table above lists the Spearman's rank correlation
  coefficient, $\rho$, and the associated false correlation
  probability, for different physically-motivated X-ray fitting parameters
  versus radio flux density.}
\vspace{-1.0\baselineskip}
\end{footnotesize}
\end{table}

\vspace{1.0in}

\begin{table}[htb!]
\caption{Rank Correlations of Selected X-ray Reflection Model Parameters}
\begin{footnotesize}
\begin{center}
\begin{tabular}{llll}
Parameter 1 & Parameter 2 & Spearman's $\rho$ & False Alarm Probability	\\
\tableline

reflection flux & disk flux & 0.881 & $3.0\times 10^{-7}$ \\

disk norm. & disk kT & -0.795 & $2.8\times 10^{-5}$ \\

power-law flux & reflection flux & -0.662 & 0.002 \\

disk kT & $\xi$ & 0.658 & 0.002 \\

reflection flux & $\Gamma$ & -0.340 & 0.142 \\

disk norm. & $R_{in}$ & 0.320 & 0.169 \\

reflection/total flux & $R_{in}$ & 0.271 & 0.248 \\

reflection/total flux & $\Gamma$ & -0.254 & 0.280 \\

reflection flux & $R_{in}$ & 0.234 & 0.321 \\

\tableline
\end{tabular}
\vspace*{\baselineskip}~\\ \end{center} 
\tablecomments{The table above lists the Spearman's rank correlation
  coefficient, $\rho$, and the associated false correlation
  probability, for different physically--motivated X-ray fitting parameters.}
\vspace{-1.0\baselineskip}
\end{footnotesize}
\end{table}

\clearpage

\centerline{~\psfig{file=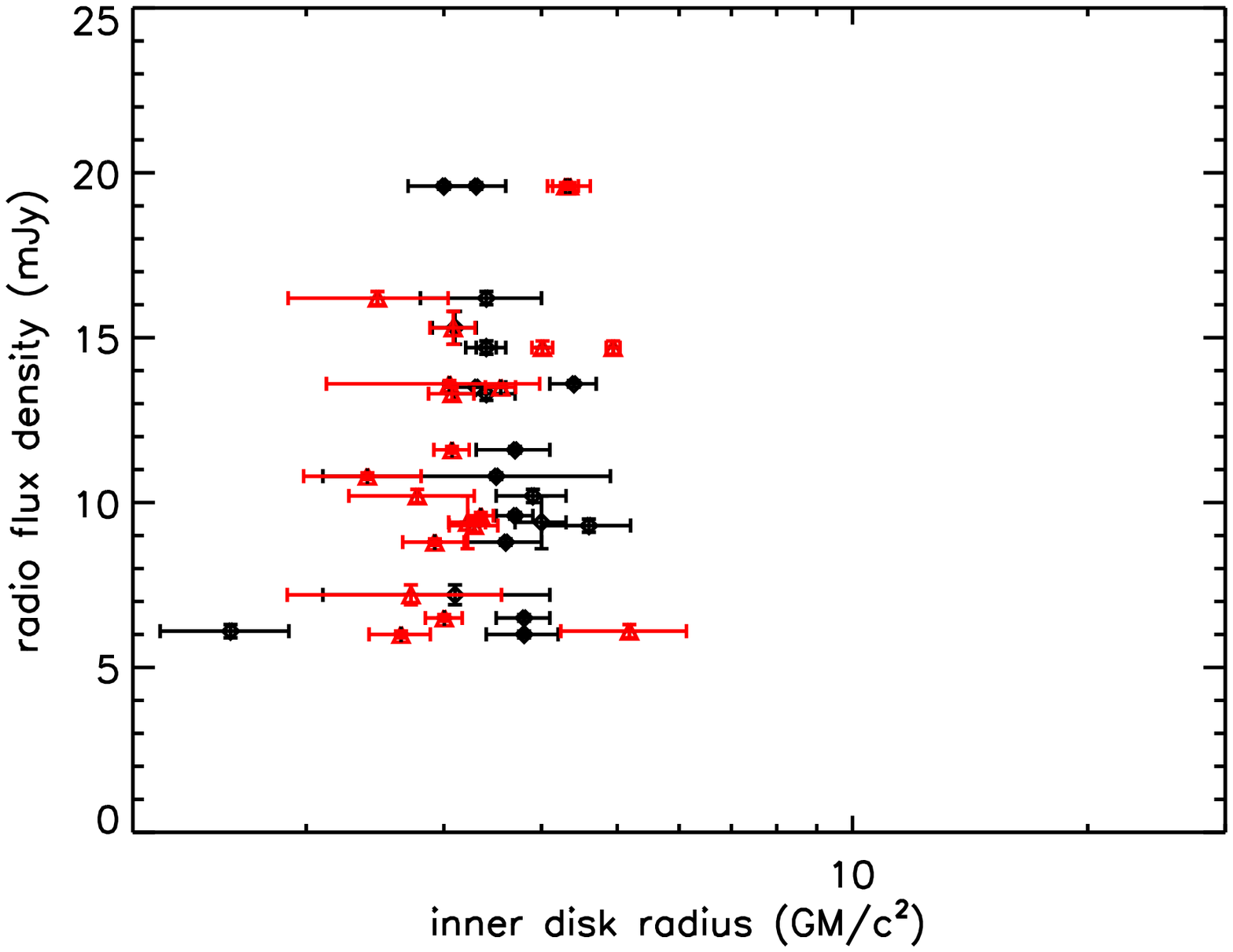,width=5.0in}~}
\figcaption[h]{\footnotesize Variations in the inner radius of the
  accretion disk are one plausible means by which the power in a
  relativistic jet might be modulated in black hole systems.  The
  figure above plots radio flux density versus the inner disk radius
  as measured by fits to the broad Fe K line with a Laor line profile
  (black) and fits to the thermal continuum with a simple disk model
  (red).  Despite strong radio variability, there is clearly little
  variability in the inner disk radius.  The Spearman's rank
  correlation coefficients with radio flux for these radius estimators
  are just $\rho = -0.369$ (line) and $\rho = 0.312$ (continuum),
  indicating that radio jet power and inner disk radius are not
  stongly correlated.}
\medskip

\centerline{~\psfig{file=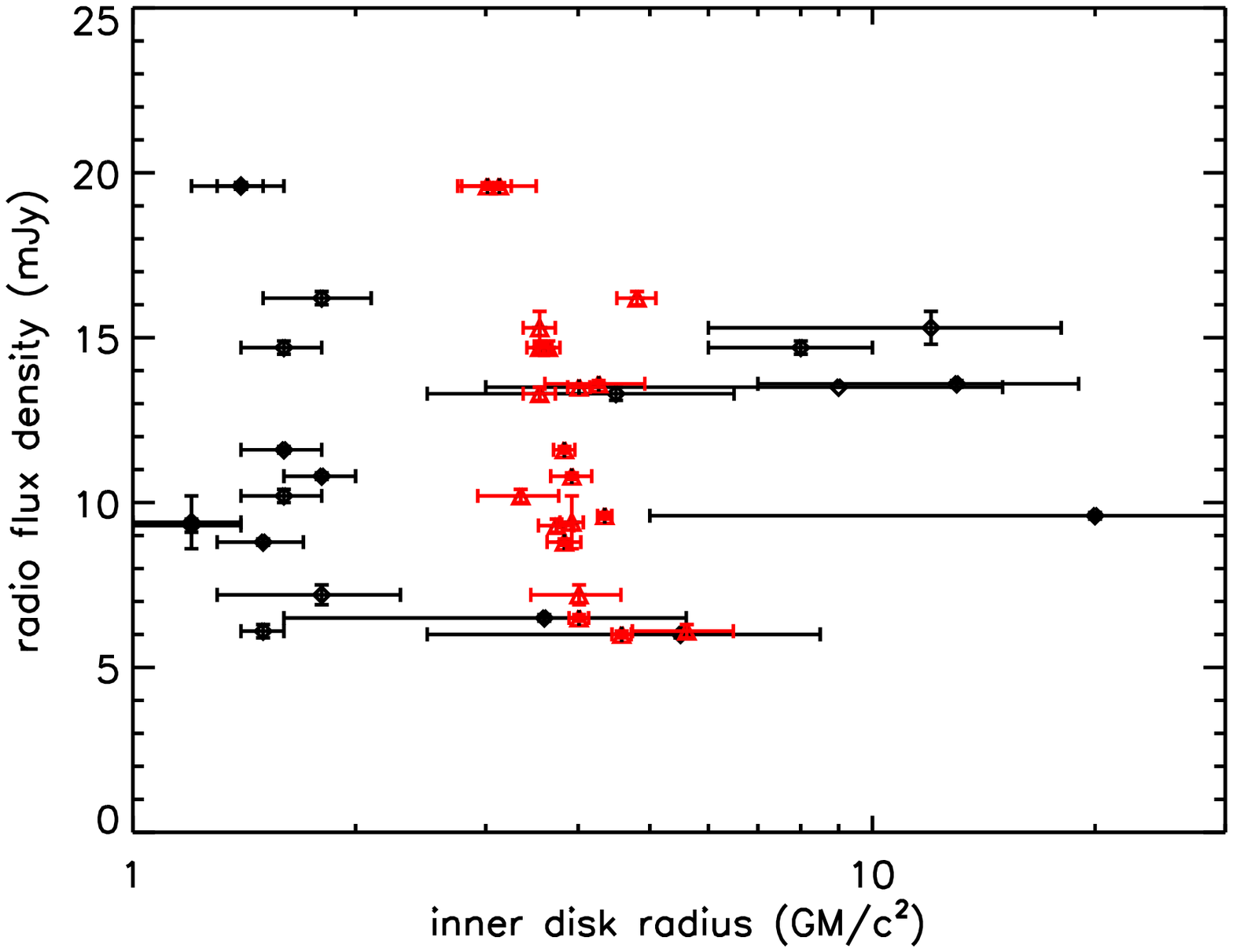,width=5.0in}~}
\figcaption[h]{\footnotesize Variations in the inner radius of the
  accretion disk are a plausible means by which the power in a
  relativistic jet might be modulated in black hole systems.  The
  figure above plots radio flux density versus the inner disk radius
  as measured by fits to the disk reflection spectrum with variable
  relativistic blurring (black) and fits tot he thermal continuum with
  a simple disk model (red).  Despite strong radio variability, there
  is clearly little variability in the inner disk radius.  The
  Spearman's rank correlation coefficients with radio flux for these
  radius estimators are just $\rho = 0.090$ (line) and $\rho = -0.533$
  (continuum).  After filtering out two suspect continuum--derived
  radii estimates (see Section 5.1), neither estimator of the inner
  disk radius is significantly correlated with the radio flux.  }
\medskip

\clearpage

\centerline{~\psfig{file=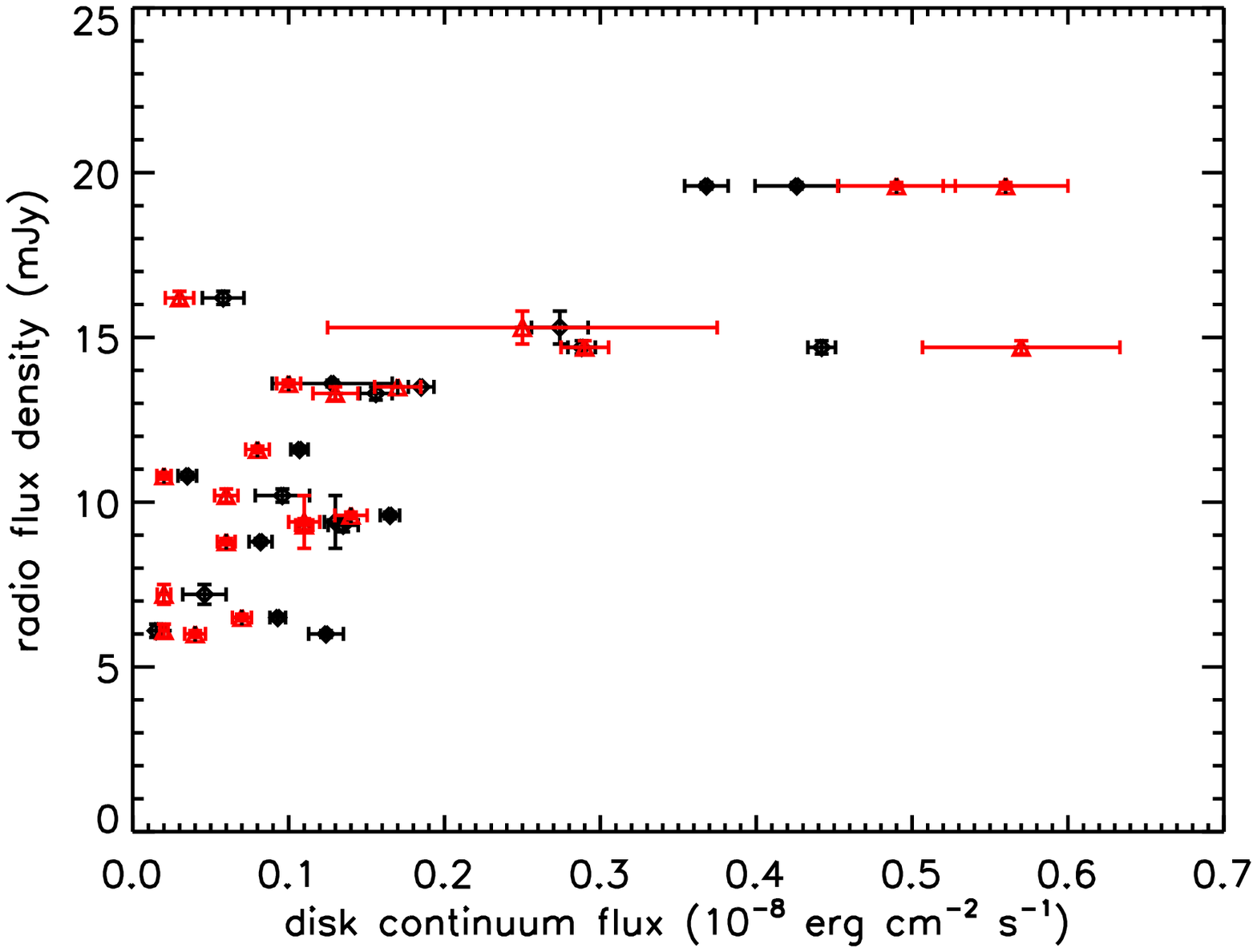,width=5.0in}~}
\figcaption[h]{\footnotesize The flux from an accretion disk --
  especially if the flux traces $\dot{m}$ directly -- may serve to
  modulate jet properties.  The plot above shows the radio flux
  density observed from Cygnus X-1, versus the unabsorbed
  0.8--10.0~keV disk component flux measured in simple fits (black)
  and blurred reflection fits (red) to {\it Suzaku} spectra.  Both
  estimators of the disk continuum flux are postively correlated with
  radio flux at the 99.7\% level of confidence.}
\medskip

\centerline{~\psfig{file=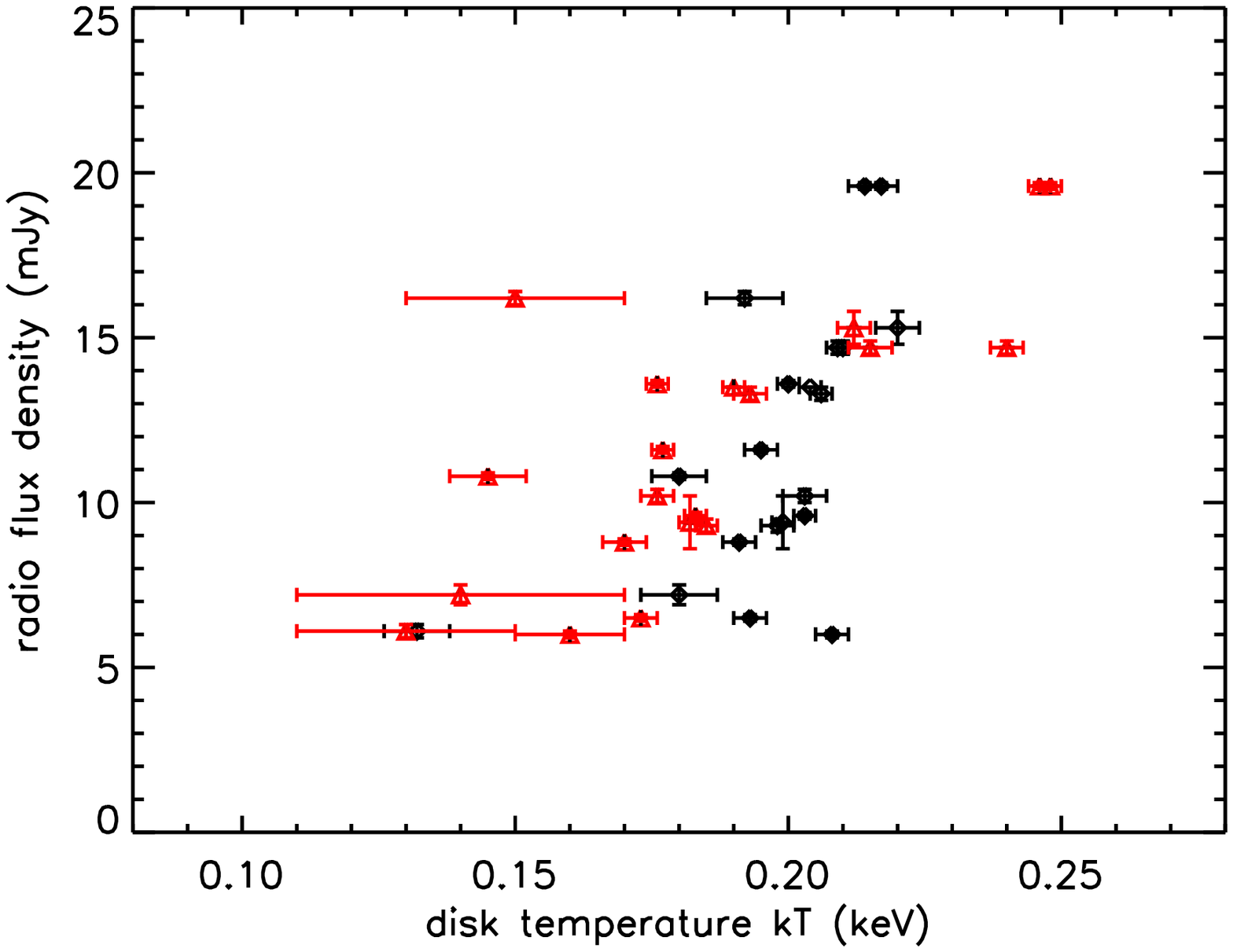,width=5.0in}~}
\figcaption[h]{\footnotesize The figure above plots radio jet flux
  density versus inner disk color temperature, based on fits to the
  disk continuum in both phenomenological (black) and blurred
  reflection (red) spectral models.  As with estimates of the disk
  continuum flux, both estimates of the inner disk color temperature
  are positively correlated with radio flux at more than 99.6\% confidence.}

\clearpage

\centerline{~\psfig{file=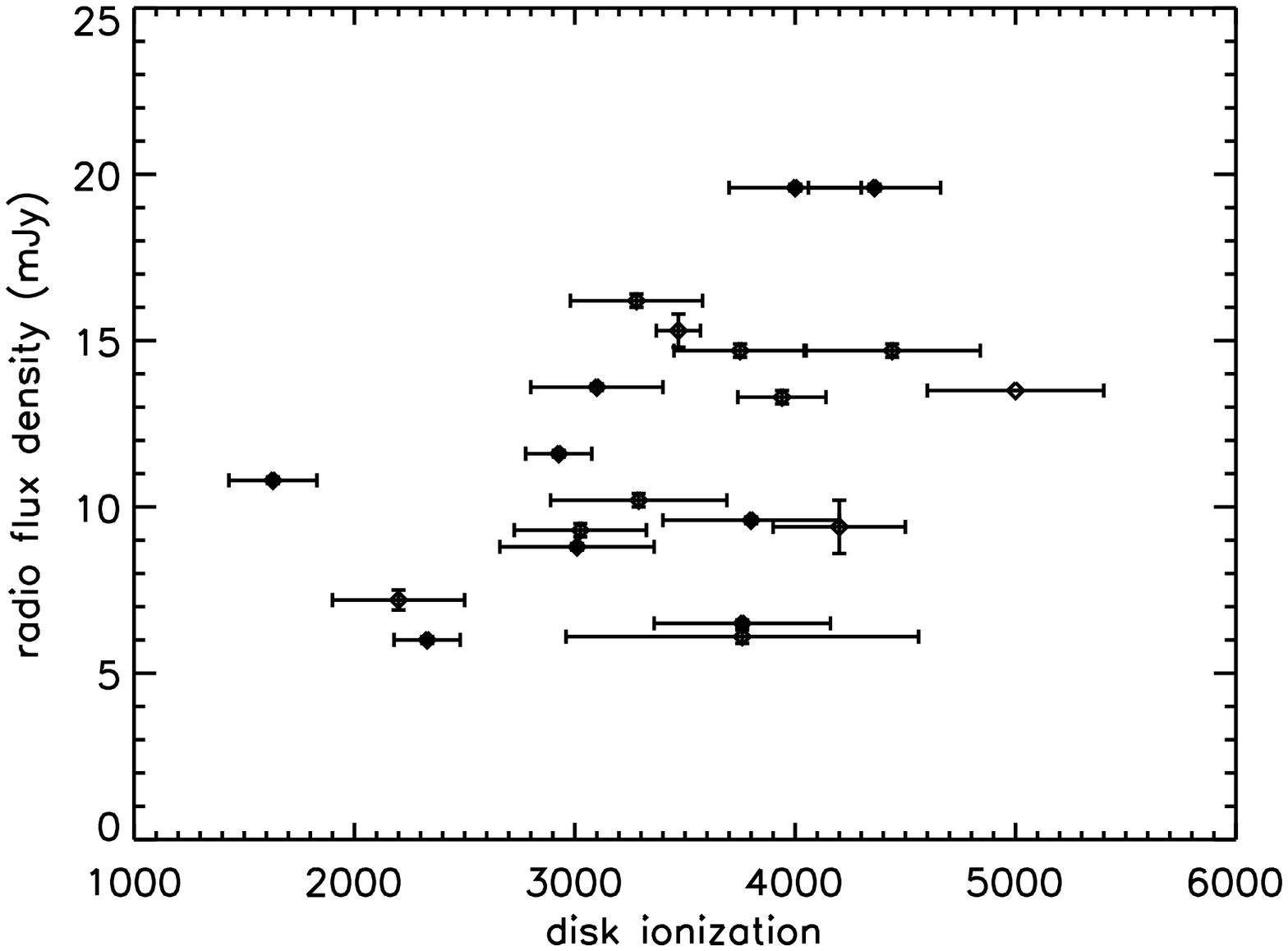,width=5.0in}~}
\figcaption[h]{\footnotesize The figure above plots radio flux density
  versus the ionization of the accretion disk, as measured in fits
  with a relativistically--blurred disk reflection model.  The general
  trend seen in this plot is supported by a Spearman's rank
  correlation test, which returns a coefficient of $\rho = 0.415$,
  (corresponding to a 6.9\% chance of a false correlation).  Though
  tentative, this correlation is as strong as that between radio flux
  and the hard X-ray count rate ($\rho = 0.455$, see Figure 5), and it
  may suggest that jet production is partly tied to disk ionization.
  A simple physical explanation might be that poloidal magnetic field
  lines are more easily anchored in the ionized skin of a disk than in
  a low-ionization disk with less free charge.}
\medskip

\centerline{~\psfig{file=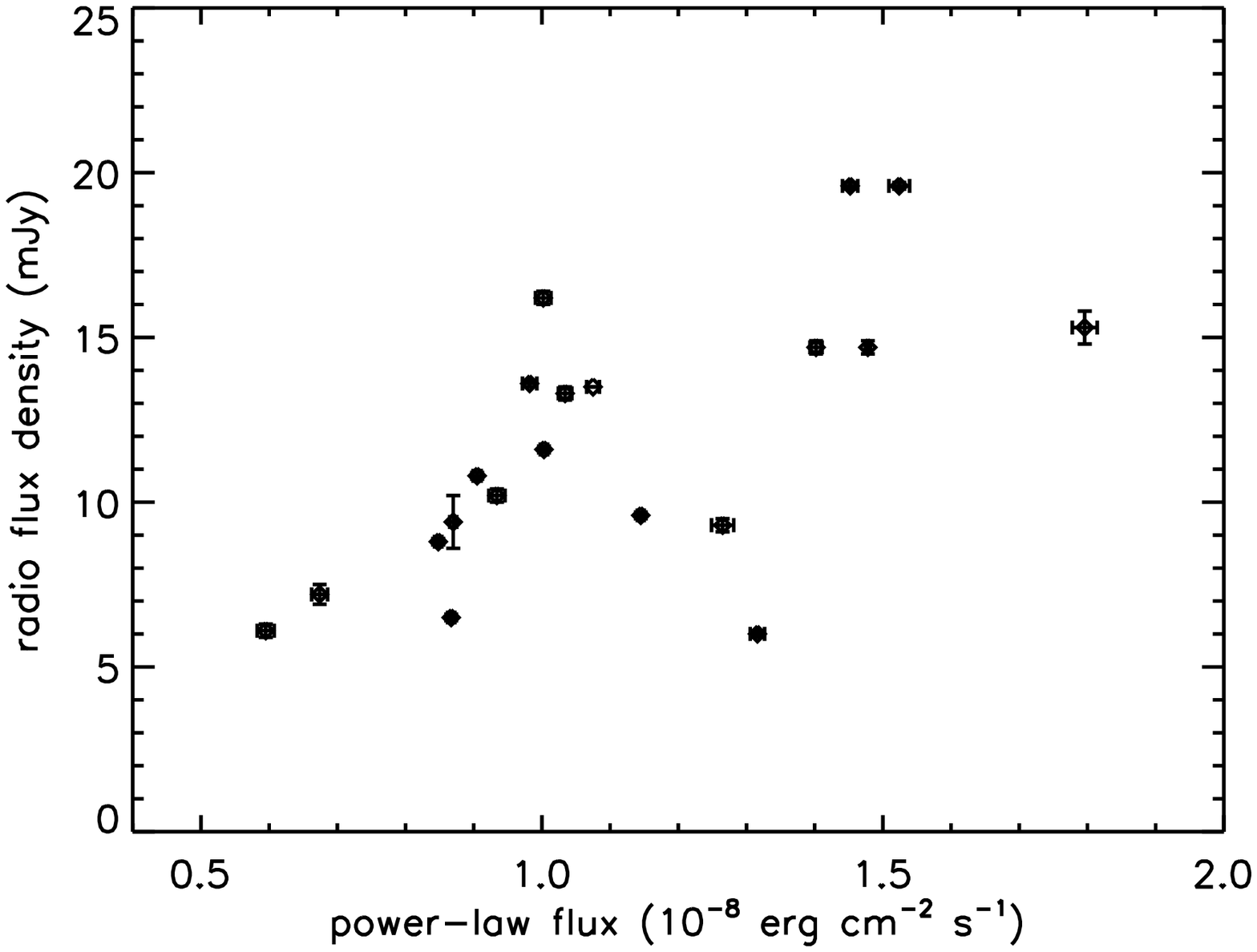,width=5.0in}~}
\figcaption[h]{\footnotesize The corona -- if traced effectively by a
  power-law component -- may also modulate jet properties in accreting
  black holes.  The plot above shows the radio flux density observed
  from Cygnus X-1, versus the unabsorbed 0.8--10.0~keV power-law
  component flux measured in phenomenological fits to {the \it Suzaku}
  spectra.  These paramters are positively correlated at the 99.8\%
  level of confidence.  At least in the disk/corona decomposition
  possible in simple spectral fits, then, both the disk and corona are
  correlated with radio jet flux, and both may contribute to jet
  formation and modulation.}
\medskip

\clearpage

\centerline{~\psfig{file=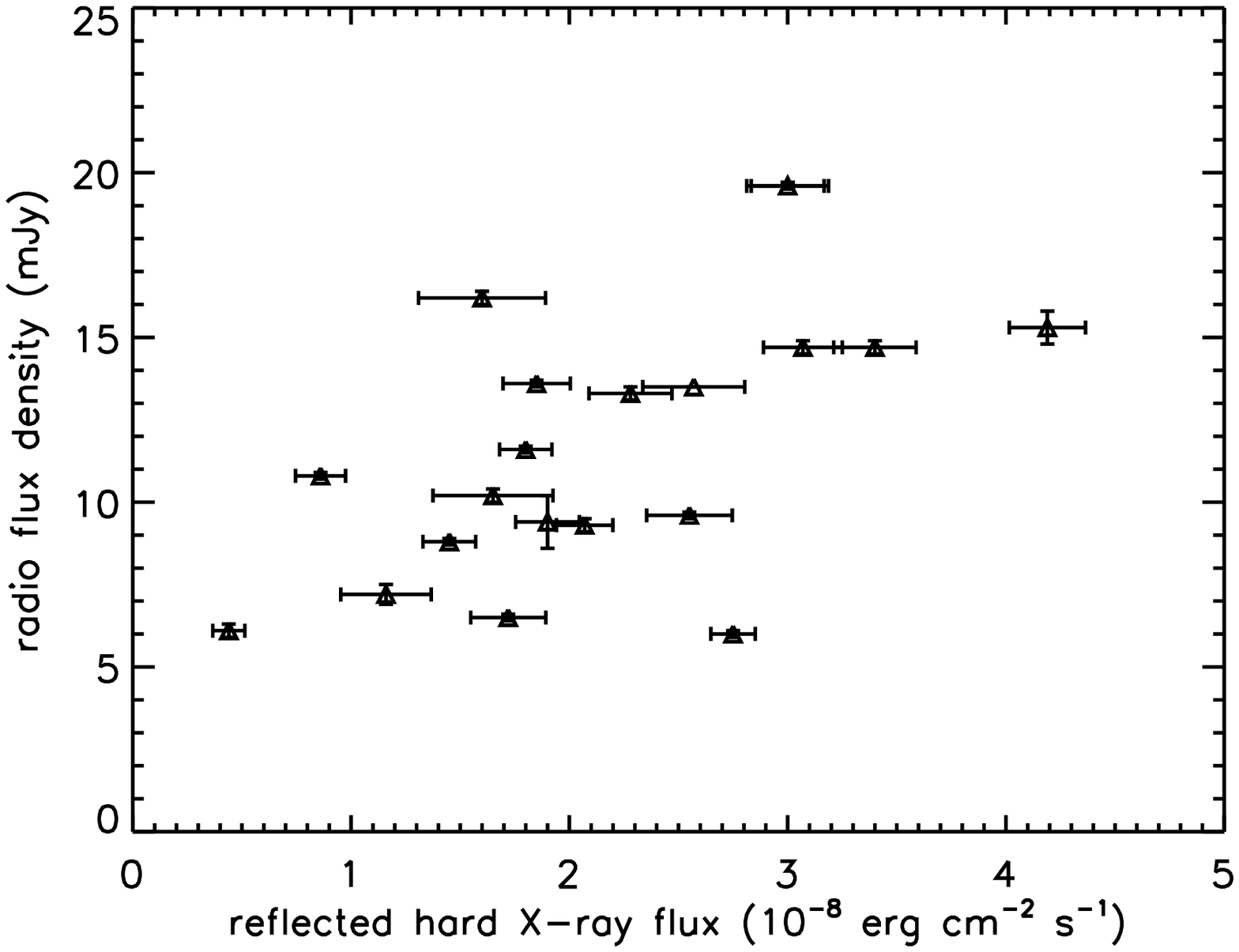,width=5.0in}~}
\figcaption[h]{\footnotesize The plot above shows the radio flux
  density observed from Cygnus X-1, versus the unabsorbed
  0.8--10.0~keV reflected disk flux measured in more physical model
  fits to the {\it Suzaku} spectra.  In this X-ray spectral
  decomposition, the radio flux is anti-correlated with the power-law
  flux (93\% confidence), and positively correlated with the reflected
  flux (99\% confidence).  This result may constitute specific support
  for a plasma ejection model of the hard X-ray corona, in which
  radiation pressure from reflected disk flux helps to launch the base
  of a jet (Beloborodov 1999).}
\medskip

\centerline{~\psfig{file=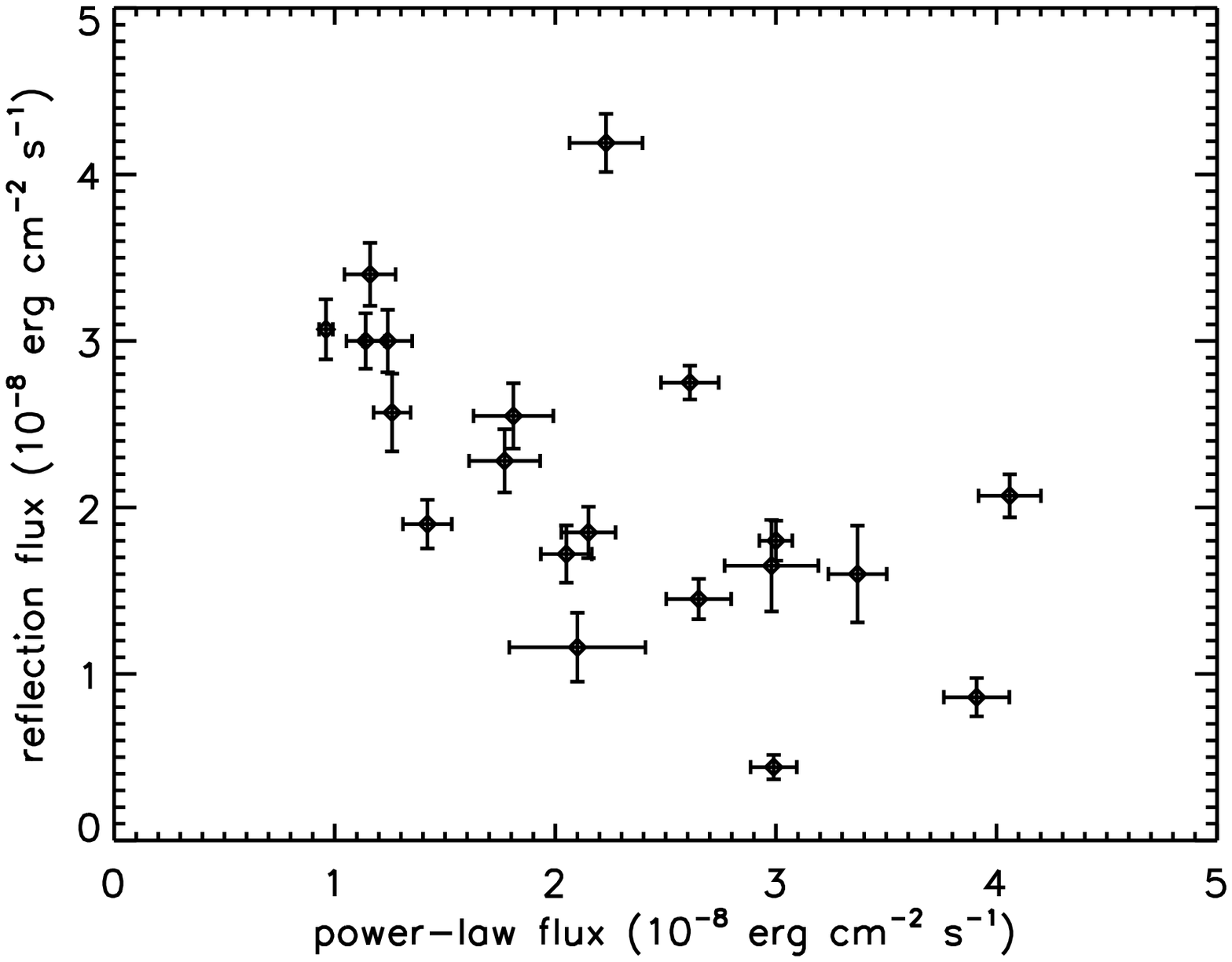,width=5.0in}~}
\figcaption[h]{\footnotesize The figure above plots the flux in the
  blurred disk reflection component, versus the flux in the power-law
  component.  Flux from the power-law component is assumed to be
  directly observed.  In simple reflection models, reflection flux
  should vary linearly with the power-law flux.  However, if the
  power-law flux originates close to a spinning black hole,
  gravitational light bending can alther the flux relationship.  The
  anti-correlation shown in the plot above ($\rho = -0.662$, giving a
  0.2\% chance of false correlation) is suggestive of a source height
  that is close to $\simeq 20~GM/c^{2}$ according to the calculations
  of Miniutti \& Fabian (2004). }
\medskip


\begin{references}

\reference{} Anders, E., \& Grevesse, N., 1989, Geochimica et
Cosmochimica Acta, 53, 197

\reference{} Arnaud, K. A., and Dorman, B., 2000, XSPEC is available
via the HEASARC on-line service, provided by NASA/GSFC

\reference{} Allen, S. W., Dunn, R. J. H., Fabian, A. C., Taylor,
G. B., \& Reynolds, C. S., 2006, MNRAS, 372, 21

\reference{} Balucinska-Church, M., Church, M. J., Charles, P. A.,
Nagase, F., LaSala, J., Barnard, R., 2000, MNRAS, 311, 861

\reference{} Bardeen, J. M., \& Petterson, J. A., 1975, ApJ, 195, L65

\reference{} Barr, P., White, N. E., \& Page, C. G., 1985, 216, 65

\reference{} Beckwith, K., \& Done, C., 2004, MNRAS, 352, 353

\reference{} Bell, M. E., et al., 2011, MNRAS, 411, 402

\reference{} Beloborodov, A., 1999, ApJ, 510, L123

\reference{} Blandford, R. D., \& Payne, D. G., 1982, MNRAS, 199, 883

\reference{} Blandford, R. D., \& Znajek, R. L., 1977, MNRAS, 179, 433

\reference{} Blum, J. L., Miller, J. M., Fabian, A. C., Miller, M. C.,
Homan, J., van der Klis, M., Cackett, E. M., \& Reis, R. C., 2009,
ApJ, 706, 60

\reference{} Blustin, A. J., Page, M. J., Fuerst, S. V.,
Branduardi-Raymont, G., Ashton, C. E., 2005, A\&A, 431, 111

\reference{} Brenneman, L., \& Reynolds, C. S., 2006, ApJ, 652, 1028

\reference{} Brocksopp, C., Tarasov, A. E., Lyuty, V. M., Roche, P., 1999, A\&A, 343, 861

\reference{} Calvet, N., Hartmann, L., \& Kenyon, S., 1993, ApJ, 402, 623

\reference{} Chartas, G., Kochanek, C. S., Dai, X., Poindexter, S., Garmire, G., 2009, ApJ, 693, 174

\reference{} Chen, B., Dai, x., Kochanek, C. S., Chartas, G.,
Blackburne, J. A., \& Kozlowski, S., 2011, ApJ, subm., arxiv:1106.6052

\reference{} Dai, X., Kochanek, C. S., Chartas, G., Kozlowski, S.,
Morgan, C. W., Garmire, G., \& Agol, E., 2010, ApJ, 709, 278

\reference{} Dovciak, M., Karas, V., \& Yaqoob, T., 2004, ApJS, 153, 205

\reference{} Duro, R., et al., 2011, A\&A, in press, arxiv:1108.1157

\reference{} Ebisawa, K., et al., 1994, PASJ, 46, 375

\reference{} Ebisawa, K., Ueda, Y., Inoue, H., Tanaka, Y., \& White, N. E., 1996, ApJ, 467, 419

\reference{} Esin, A. A., McClintock, J. E., \& Narayan, R., 1997, ApJ, 489, 865

\reference{} Fabian, A. C., et al., 2009, Nature, 459, 540

\reference{} Fabian, A. C., et al., 2012, MNRAS, in press

\reference{} Falcke, H., Kording, E., \& Markoff, S., 2004, A\&A, 414, 895

\reference{} Fender, R., Gallo, E., Russell, D., 2010, MNRAS, 406, 1425

\reference{} Frank, J., King, A., \& Raine, D., 2002, in ``Accretion
Power in Astrophyiscs'', Cambridge University Press, Cambridge

\reference{} Frontera, F., et al., 2001, ApJ, 546, 1027

\reference{} Gallo, E., Fender, R., \& Pooley, G., 2003, MNRAS, 344, 60

\reference{} George, I. M., \& Fabian, A. C., 1991, MNRAS, 249, 352

\reference{} Gierlinski, M., Done, C., \& Page, K., 2009, MNRAS, 392, 1106

\reference{} Gies, D. R., et al., 2008, ApJ, 678, 1237

\reference{} Gilfanov, M., Churazov, E., \& Revnivtsev, M., 1999, A\&A, 352, 182

\reference{} Gou, L., et al., 2011, ApJ, submitted, arxiv:1106.3690

\reference{} Gultekin, K., Cackett, E. M., Miller, J. M., Di Matteo,
T., Markoff, S., \& Richstone, D. O., 2009, ApJ, 706, 404

\reference{} Hanke, M., Wilms, J., Nowak, M. A., Pottschmidt, K.,
Schulz, N. S., Lee, J. C., 2009, ApJ, 630 330

\reference{} Jones, S., McHardy, I., Moss, D., Seymour, N., Breedt,
E., Uttley, P., Kording, E., Tudose, V., 2011, MNRAS, 412, 2641
 
\reference{} Junor, W., Biretta, J. A., \& Livio, M., 1999, Nature, 401, 891

\reference{} King, A. L., et al., 2011, ApJ, 729, 19

\reference{} Kraemer, S. B., et al., 2006, ApJS, 167, 161

\reference{} Kubota, A., et al., 2007, PASJ, 59, 185

\reference{} Livio, M., Pringle, J., \& King, A. R., 2003, ApJ, 593, 184

\reference{} Maccarone, T., 2002, MNRAS, 336, 1371

\reference{} Magdziarz, P., \& Zdziarski, A. A., 1995, MNRAS, 273, 837

\reference{} Maitra, D., Miller, J. M., Markoff, S., \& King, A. L., 2011, ApJ, 735, 107

\reference{} Makishima, K., et al., 2008, PASJ, 60, 585

\reference{} Markoff, S., Falcke, H., \& Fender, R., 2001, A\&A, 372, L25

\reference{} Markoff, S., Nowak, M., \& Wilms, J., 2005, ApJ, 635, 1203

\reference{} Marscher, A. P., et al., 2008, Nature, 452, 966

\reference{} Mauche, C., \& Raymond, J. C., 2000, ApJ, 541, 924

\reference{} McClintock, J. E., \& Remillard, R. A., 2005, to appear
in ``Compact Stellar X-ray Sources,'' eds. W. H. G. Lewin and M. van
der Klis, Cambridge: Cambridge University press, astro-ph/0306213

\reference{} McClintock, J. E., Narayan, R., Gou, L., Liu, J., Penna,
R. F., \& Steiner, J. F., 2010, AIPC, 1248, 101

\reference{} McNamara, B. R., Kazemzadeh, F., Rafferty, D. A., Birzan,
L., Nulsen, P. E. J., Kirkpatrick, C. C., \& Wise, M., W., 2009, ApJ,
689, 594

\reference{} Merloni, A., Ross, R. \& Fabian, A. C., 2000, MNRAS, 313, 193

\reference{} Merloni, A., Heinz, S., Di Matteo, T., 2003, MNRAS, 345, 1057

\reference{} Miller, J. M., 2007, ARA\&A, 45, 441

\reference{} Miller, J. M., Cackett, E. M., Reis, R. C., 2009, ApJ, 707, L77

\reference{} Miller, J. M., Wojdowski, P., Schulz, N. S., Marshall,
H. L., Fabian, A. C., Remillard, R. A., Wijnands, R., \& Lewin,
W. H. G., 2005, ApJ, 620, 398

\reference{} Miller, J. M., Reynolds, C. S., Fabian, A. C., Miniutti,
G., \& Gallo, L. C., 2009, ApJ, 697 900

\reference{} Miller, J. M., et al., 2010, ApJ, 724, 1441

\reference{} Miller, J. M., Miller, M. C., \& Reyonlds, C. S., 2011, ApJ, 731, L5

\reference{} Mineshige, S., Hirano, A., Kitamoto, S., Yamada, T., Fukue, J., 1994, ApJ, 426, 308

\reference{} Miniutti, G., \& Fabian, A. C., 2004, MNRAS, 349, 1435

\reference{} Miniutti, G., Fabian, A. C., \& Miller, J. M., 2004, MNRAS, 351, 466

\reference{} Miniutti, G., et al., 2007, PASJ, 59, S315

\reference{} Mitsuda, K., et al., 1984, PASJ, 36, 741

\reference{} Nandra, K., O'Neill, P. M., George, I. M., \& Reeves, J. N., 2007, MNRAS, 382, 194

\reference{} Narayan, R., \& McClintock, J. E., 2012, MNRAS, 419, L69

\reference{} Nayakshin, S., \& Kallman, T. R., 2001, ApJ, 546, 406

\reference{} Nowak, M. A., 2009, scripts available on-line via space.mit.edu/CXC/software/suzaku/index.html

\reference{} Nowak, M. A., et al., 2011, ApJ, 728, 13

\reference{} Orosz, J. A., McClintock, J. E., Aufdenberg, J. P.,
Remillard, R. A., Reid, M. J., Narayan, R., Gou, L., 2011, ApJ, in
press, arxiv:1106.3689

\reference{} Park, S. Q., et al., 2004, ApJ, 610, 378

\reference{} Pooley, G. G., \& Fender, R., 1997, MNRAS, 292, 925

\reference{} Pooley, G. G., Fender, R., \& Brocksopp, C., 1999, MNRAS, 302, L1

\reference{} Reid, M. J., McClintock, J. E., Narayan, R., Gou, L.,
Remillard, R. A., \& Orosz, J. A., 2011, ApJ, 742, 83

\reference{} Reis, R. C., Fabian, A. C., \& Miller, J. M., 2010, MNRAS, 402, 836

\reference{} Reis, R. C., Fabian, A. C., Ross, R. R., Miniutti, G.,
Miller, J. M., \& Reynolds, C. S., 2008, MNRAS, 387, 1489

\reference{} Reynolds, C. S., Garofalo, D., \& Begelman, M. C., 2006, ApJ, 651, 1023

\reference{} Reynolds, C. S., \& Wilms, J., 2000, ApJ, 533, 821

\reference{} Reynolds, M. T., Miller, J. M., Homan, J., \& Miniutti, G., 2010, ApJ, 709, 358

\reference{} Reynolds, M. T., \& Miller, J. M., 2012, ApJ, submitted, arxiv:1112.2249

\reference{} Ross, R. R., \& Fabian, A. C., 2005, MNRAS, 358, 211

\reference{} Rossi, S., Homan, J., Miller, J. M., \& Belloni, T.,
2005, MNRAS, 360, 763

\reference{} Russell, D. M., Miller-Jones, J. C. A., Maccarone, T. J.,
Yang, Y. J., Fender, R. P., \& Lewis, F.,, 2011, ApJ, 739, L19

\reference{} Rykoff, E. S., Miller, J. M., Steeghs, D., \& Torres, M. A. P., 2007, ApJ, 666, 1129

\reference{} Schulz, N. S., Cui, W., Canizares, C. R., Marshall,
H. L., Lee, J. C., Miller, J. M., Lewin, W. H. G., 2002, ApJ, 565,
1141

\reference{} Shakura, N. I., \& Sunyaev, R. A., 1973, A\& A, 86, 121

\reference{} Shimura, T., \& Takahara, F., 1995, ApJ, 445, 780

\reference{} Sikora, M., Stawarz, L., Lasota, J.-P., 2007, ApJ, 658, 815

\reference{} Sobczak, G. J., McClintock, J. E., Remillard, R. A., Cui,
W., Levine, A. M., Morgan, E. H., Orosz, J. A., Bailyn, C. D., 2000,
ApJ, 544, 993

\reference{} Sowers, J. W., Gies, D. R., Bagnuolo, W. G., Shafter,
A. W., Wiemker, R., Wiggs, M. S., 1998, ApJ, 506, 424

\reference{} Steiner, J. F., McClintock, J. E., Remillard, R. A., Gou,
L., Yamada, S., \& Narayan, R., 2010, ApJ, 718, L117

\reference{} Steiner, J. F., et al., 2011, MNRAS, in press, arxiv:1010.1013

\reference{} Stirling, A. M., Spencer, R. E., de la Force, C. J.,
Garrett, M. A., Fender, R. P., Ogley, R. N., 2001, MNRAS, 327, 1273

\reference{} Tanaka, Y., et al., 1995, Nature, 375, 659

\reference{} Tomsick, J. A., Yamaoka, K., Corbel, S., Kaaret, P.,
Kalemci, E., Migliari, S., 2009, ApJ, 707, L87

\reference{} Torii, S., Yamada, S., Makishima, K., Sakurai, S.,
Nakazawa, K., Noda, H., Done, C., Takahashi, H., Gandhi, P., 2011,
PASJ, 63, 771

\reference{} Uttley, P., Wilkinson, T., Cassatella, P., Wilms, J.,
Pottschmidt, K., Hanke, M., Bock, M., 2011, MNRAS, 404, L60

\reference{} Vaughan, S., \& Fabian, A. C., 2004, MNRAS, 348, 1415

\reference{} Wilkins, D. R., \& Fabian, A. C., 2011, MNRAS, 414, 1269

\reference{} Wilkinson, T., \& Uttley, P., 2009, MNRAS, 397, 666

\reference{} Wilms, J., Allen, A., McCray, R., 2000, ApJ, 542, 914

\reference{} Wilms, J., Nowak, M. A., Pottschmidt, K., Pooley, G. G., \& Fritz, S., 2006, A\&A, 447, 245

\reference{} Zdziarski, A. A., Lubinski, P., Gilfanov, M., Revnivtsev, M., 2003, MNRAS, 342, 355

\reference{} Zdziarski, A. A., Skinner, G. K., Pooley, G. G.,
Lubinski, P., 2011, MNRAS, in press, arxiv:1103.2988

\reference{} Zimmerman, E. R., Narayan, R., McClintock, J. E., \&
Miller, J. M., 2005, ApJ, 618, 832

\reference{} Zwart, J. T. L., et al., 2008, MNRAS, 391, 1545

\end{references}
\end{document}